\documentclass[usenatbib,usegraphicx]{mn2e}
\usepackage{amsmath}
\usepackage{amssymb}
\usepackage{mathrsfs}
\usepackage{rotating}

\newcommand{\rmd}{{\mathrm{d}}}
\newcommand{\Msun}{\:{\rm M}_{\odot}}
\newcommand{\gammao}{\gamma_{_{\cal{E}}}}
\newcommand{\Mtwohun}{{\rm M}_{\rm halo}}
\newcommand{\rhalf}{{\rm r}_{_{1/2}}}
\newcommand{\Rhalf}{{\rm R}_{_{\rm e}}}
\newcommand{\Reff}{{\rm R}_{_{\rm e}}}
\newcommand{\rcore}{{\rm r}_{_{\rm core}}}
\newcommand{\rlimit}{{\rm r}_{_{\rm lim}}}
\newcommand{\rcut}{{\rm r}_{_{\rm cut}}}
\newcommand{\rmax}{{\rm r}_{_{\rm high}}}
\newcommand{\rbeta}{{\rm r}_{_{\beta}}}
\newcommand{\rzero}{{\rm R}_{_{0}}}
\newcommand{\req}{{\rm r}_{_{\rm eq}}}
\newcommand{\rthree}{{\rm r}_{_3}}
\newcommand{\rsig}{r_\sigma}
\newcommand{\Mt}{{\rm M}_{\rm t}}
\newcommand{\rg}{{\rm r}_{\rm g}}
\newcommand{\Mhalf}{{\rm M}_{_{1/2}}}
\newcommand{\Lhalf}{{\rm L}_{_{1/2}}}
\newcommand{\moverl}{\Upsilon^{{\rm I}}_{_{1/2}}}
\newcommand{\moverlI}{\Upsilon^{{\rm I}}_{_{1/2}}}
\newcommand{\sigmalos}{\sigma_{_{\rm{los}}}}
\newcommand{\sigmatot}{\sigma_{_{\rm{tot}}}}
\newcommand{\Lsun}{{\rm L}_\odot}
\newcommand{\LsunV}{{\rm L}_{\odot,{\rm V}}}
\newcommand{\LV}{{\rm L}_{\rm V}}

\newcommand{\LI}{{\rm L}_{\rm I}}
\newcommand{\kms}{{\rm km \: s}^{-1}}
\newcommand{\LCDM}{\Lambda{\rm CDM}}
\newcommand{\beq}{\begin{equation}}
\newcommand{\eeq}{\end{equation}}

\newcommand{\data}{{\mathscr{V}}}
\newcommand{\dataD}{{\mathscr{D}}}
\newcommand{\Einastop}{{\mathscr{M}}}

\newcommand{\ave}[1]{\langle #1 \rangle}
\newcommand{\avelos}{\langle \sigma_{_{\rm los}}^2 \rangle}
\newcommand{\avelosap}{\avelos_{_{\Rhalf}}}

\begin{document}

\title[Accurate masses for spheroidal galaxies]{Accurate masses for dispersion-supported galaxies}

\author[J. Wolf et al.]
{Joe Wolf$^{1^\star}$,
Gregory D. Martinez$^1$,
James S. Bullock$^1$,
Manoj Kaplinghat$^1$, \newauthor
Marla Geha$^2$,
Ricardo R. Mu{\~n}oz$^2$,
Joshua D. Simon$^3$,
Frank F. Avedo$^1$\\
$^1$Center for Cosmology, Department of Physics and Astronomy, University of California, Irvine, CA 92697\\
$^2$Astronomy Department, Yale University, New Haven, CT 06520\\
$^3$Observatories of the Carnegie Institution of Washington, Pasadena, CA 91101\\
$^\star$wolfj@uci.edu\\
}

\date{\today}

\maketitle

\begin{abstract}
We derive an accurate mass estimator for dispersion-supported stellar
systems and demonstrate its validity by analyzing resolved
line-of-sight velocity data for globular clusters, dwarf galaxies,
and elliptical galaxies. Specifically, by manipulating the spherical
Jeans equation we show that the mass enclosed within the 
3D deprojected half-light radius $\rhalf$ can be determined with only mild 
assumptions about the spatial variation of the stellar velocity dispersion anisotropy
as long as the projected velocity dispersion profile is fairly flat near the 
half-light radius, as is typically observed. We find $\Mhalf = 3 \, G^{-1} \, 
\avelos \, \rhalf \simeq 4 \, G^{-1} \, \avelos \, \Rhalf$, where
$\avelos$ is the luminosity-weighted square of the 
line-of-sight velocity dispersion and $\Rhalf$ is the 2D projected
half-light radius. While deceptively familiar in form, this formula is
not the virial theorem, which cannot be used to determine accurate
masses unless the radial profile of the total mass is known {\em a priori}.
We utilize this finding to show that all of the Milky Way dwarf
spheroidal galaxies (MW dSphs) are consistent with having formed 
within a halo of mass approximately $3 \times 10^9 \Msun$, assuming a
$\LCDM$ cosmology. The faintest MW dSphs seem to have formed in dark
matter halos that are at least as massive as those of the brightest
MW dSphs, despite the almost five orders of magnitude spread in luminosity
between them. We expand our analysis to the full range of observed
dispersion-supported stellar systems and examine their dynamical I-band
mass-to-light ratios $\moverlI$. 
The $\moverlI$ vs. $\Mhalf$ relation for dispersion-supported galaxies
follows a U-shape, with a broad minimum near $\moverlI \simeq 3$ that
spans dwarf elliptical galaxies to normal ellipticals, a steep
rise to $\moverlI \simeq$ 3,200 for ultra-faint dSphs, and a more
shallow rise to $\moverlI \simeq 800$ for galaxy cluster spheroids.  
\end{abstract}

\begin{keywords}
Galactic dynamics, dwarf galaxies, elliptical galaxies, galaxy formation, dark matter
\end{keywords}

\section{Introduction}
\label{section:intro}

Mass determinations for dispersion-supported galaxies based on only line-of-sight velocity measurements suffer from a notorious uncertainty associated with not knowing the intrinsic 3D velocity dispersion. The difference between radial and tangential velocity dispersions is usually quantified by the stellar velocity dispersion anisotropy, $\beta$. Many questions in galaxy formation are affected by our ignorance of $\beta$, including our ability to quantify the amount of dark matter in the outer parts of elliptical galaxies \citep{Romanowsky_03, Dekel_05}, to measure the mass profile of the Milky Way from stellar halo kinematics \citep{Battaglia_05, Dehnen_06}, and to infer accurate mass distributions in dwarf spheroidal galaxies (dSphs) \citep{Gilmore_07, Strigari_07b}.

Here we use the spherical Jeans equation to show that for each
dispersion-supported galaxy, there exists one radius within which the
integrated mass as inferred from the line-of-sight velocity dispersion
is largely insensitive to $\beta$, and that this radius is approximately 
equal to $\rthree$, the location where the log-slope of the 3D tracer density 
profile\footnote{In this paper we will often refer to the stellar number density 
profile, but this work is applicable to any tracer system, including 
planetary nebulae and globular clusters that trace galaxy potentials, 
and galaxies that trace galaxy cluster potentials.} is $-3$ (i.e., 
$\rmd \ln n_\star / \rmd \ln r=-3$).
Moreover, the mass within $\rthree$ is well characterized by a simple
formula that depends only on quantities that may be inferred from
observations:
\beq
\label{eq:main2}
M(\rthree)= 3\, G^{-1} \, \avelos\, \rthree \, ,
\eeq
where $M(r)$ is the mass enclosed within a sphere of radius $r$, $\sigmalos$
is the line-of-sight velocity dispersion, and the brackets indicate 
a luminosity-weighted average. For a wide range of stellar light
distributions that describe dispersion-supported galaxies, $\rthree$ is
close to the 3D deprojected half-light radius $\rhalf$ and therefore
we may also write:  
\begin{eqnarray}
\label{eq:main}
\Mhalf & \equiv& M(\rhalf) \simeq 3 \, G^{-1} \, \avelos\, \rhalf \,, \\
& \simeq & 4 \, G^{-1} \, \avelos \, \Rhalf \,,\nonumber\\
& \simeq & 930 \: \left(\frac{\avelos}{\rm km^2 \, s^{-2}} \right)
 \: \left(\frac{\Rhalf}{\rm pc}\right) \: \Msun\,. \nonumber 
\end{eqnarray}
In the second line we have used $\Rhalf \simeq (3/4) \,
\rhalf$ for the 2D projected half-light radius. This approximation is
accurate to better than 2\% for exponential, Gaussian, King, Plummer,
and S\'ersic profiles (see Appendix \ref{ap:conversions} for useful
fitting formulae). 

As we show below, Equation \ref{eq:main} can be understood under the assumption
that the observed stellar velocity dispersion profile is relatively flat near $\Rhalf$.  
Clearly, one can write down self-consistent models that violate this assumption.  In these cases,
the mass uncertainty is minimized at a radius other than $\rhalf$, and Equation \ref{eq:main}
will no longer be as accurate. However,
the velocity dispersions of real galaxies in the Universe (including those we consider below)
do appear to be rather flat near the half-light radius, thus validating the use of Equation \ref{eq:main}.

In the next section we discuss  the spherical Jeans equation and our
method for determining generalized, maximum-likelihood mass profile
solutions based on line-of-sight velocity measurements.  As a point of
comparison we also discuss the virial theorem as a mass estimator for
spherical systems. In \S 3 we derive Equation \ref{eq:main}, show that it works using
real galaxy data, and explain why the $\beta$ uncertainty is minimized
at $r \simeq \rthree \simeq \rhalf$ for line-of-sight kinematics. In \S 4
we present two examples of how $\Mhalf$ determinations can be used to
inform models of galaxy formation: first, we show that the $\Mhalf$
vs. $\rhalf$ relationship for Milky Way dSph galaxies provides an
important constraint on the type of dark matter halos they were born
within; and second, we examine the dynamical half-light mass-to-light ratios for
the full range of dispersion-supported stellar systems in the
Universe and argue that this relationship can be used to inform models
of feedback. We conclude in \S 5.   

In this paper the symbol $R$ will always refer to a projected,
two-dimensional (2D) radius and the symbol $r$ will refer to a
deprojected, three-dimensional (3D) radius. 

%
\section{Review and Methodology}
\label{sec:current}

In what follows we review the virial theorem as a mass estimator for spherical systems, introduce the Jeans equation, and present our
numerical methodology for using the Jeans equation to provide general mass likelihood solutions based on line-of-sight kinematic data.  
We will use these generalized mass solutions to evaluate our $\Mhalf$ estimator in \S 3.

\subsection{The Scalar Virial Theorem }
\label{subsec:SVT}

The scalar virial theorem (SVT) is perhaps the most popular equation
used to provide rough mass constraints for spheroidal galaxies
\citep[e.g.,][]{Poveda_58, TullyFisher_77, Busarello_97}. Consider
a spherically symmetric dispersion-supported galaxy with a total
gravitating mass profile $M(r)$, which includes a 3D stellar mass density 
$\rho_\star(r) \equiv m_\star(r) \, n_\star(r)$ that truncates at a radius 
$\rlimit$.\footnote{The total mass density need not truncate at this radius.}
$m_\star(r)$ quantifies the distribution of stellar mass per normalized number
while the stellar number density $n_\star(r)$ is normalized to integrate to
unity over the stellar volume. If $m_\star(r)$ is assumed to be constant, then 
the SVT can be expressed as: 
\begin{eqnarray}
\label{eq:vt}
4 \pi G \int_0^{{\rm r}_{\rm lim}} n_\star(r) \, M(r) \, r \, dr & = &
\int_V n_\star(r) \, \sigmatot^2(r) \, \rmd^3 r \\ \nonumber
& = & \ave{\sigmatot^2} \; \: = 3 \, \avelos.
\end{eqnarray}
Note that the luminosity-weighted average of the square of the total velocity 
dispersion $\sigmatot$ is independent of $\beta$, 
and thus if one knows the number density (either by recording the position 
of every single star, or by making an assumption about how the observed 
surface brightness relates to the number density), the SVT 
provides an observationally-applicable constraint on the integrated mass 
profile within the stellar extent of the system. 

Unfortunately, the constraint associated with the SVT is
not particularly powerful as it allows a family of acceptable
solutions for $M(r)$. This point was emphasized by
\citet[][Appendix A]{Merritt_87}, who considered two extreme
possibilities for $M(r)$ (a point mass and a constant density
distribution) to show that the SVT constrains the total
mass $\Mt$ within the stellar extent $\rlimit$ to obey 
\begin{equation}
\label{eq:vtconst}
\frac{\avelos}{\ave{r_\star^{-1}}} \; \le \frac{G \, \Mt}{3} \le
\; \frac{\rlimit^3 \avelos}{\ave{r_\star^2}},
\end{equation}
where $\ave{r_\star^{-1}}$ and $\ave{r_\star^2}$ are moments of the stellar distribution.
The associated constraint is quite weak. For example, if we assume
$n_\star(r)$ follows a \citet{King_62} profile with $\rlimit / \rzero
= 5$ (typical for Local Group dwarf spheroidal galaxies) Equation
\ref{eq:vtconst} allows a large uncertainty in the mass within the
stellar extent: $0.7 \avelos \leq G \, \Mt / \rlimit \leq 20 \avelos$. 

Another common way to express the SVT is to first define a gravitational radius 
$\rg \equiv G \, \Mt^2 / |W|$ \citep{BT08}, where $W$ is the potential energy, which depends
on the unknown mass profile.
By absorbing our ignorance of the mass profile into $\rg$, we can write
the total mass as
\beq
\label{eq:svt}
\Mt = G^{-1} \, \ave{\sigmatot^2} \, \rg = 3 \, G^{-1} \, \avelos \, \rg.
\eeq
In the literature it is common to rewrite Equation \ref{eq:svt} as
\beq
\Mt = k \, G^{-1} \, \avelos \, \Rhalf,
\eeq
where $k = 3 \, \rg / \Rhalf$ is referred to as the `virial coefficient'.  
If one wishes to re-express this version of the SVT in a form 
analogous to what we have in 
Equation \ref{eq:main}, we need to relate $\Mt$ to the mass enclosed within $\rhalf$, which 
again requires knowledge of the mass profile $M(r) = f(r) \, \Mt$:
\begin{eqnarray}
M(\rhalf) &=& f(\rhalf) \, \Mt = \, f(\rhalf) \, k \, G^{-1} \, \avelos \, \rhalf \\ \nonumber
               &=& c(\rhalf) \, G^{-1} \, \avelos \, \Rhalf.
\end{eqnarray}
Note that the value of $c(\rhalf)$ depends on the (unknown) mass profile through 
{\em both} $f(\rhalf)$ and $k$.  Below, using an alternative analysis, we show that 
$c(\rhalf) = 4$ under circumstances that are fairly general for observed galaxies.  

\subsection{The Spherical Jeans Equation}
\label{subsec:JE}

Given the relative weakness of the SVT as a mass estimator, the spherical Jeans equation provides an attractive alternative.
It relates the total gravitating potential $\Phi(r)$ of a spherically symmetric, dispersion-supported, collisionless stationary system 
to its tracer velocity dispersion and tracer number density, under the assumption of dynamical equilibrium with no streaming motions:
\begin{equation}
\label{eq:jeans}
- n_{\star} \frac{\rmd \Phi}{\rmd r} = \frac{\rmd(n_{\star} \sigma_r^2)}{\rmd r} + 2 \frac{\beta \: n_{\star} \sigma_r^2}{r}.
\end{equation}
Here $\sigma_r(r)$ is the radial velocity dispersion of the stars/tracers and $\beta(r) \equiv 1- \sigma_t^2 / \sigma_r^2$ is a measure of the velocity anisotropy, where the tangential velocity dispersion $\sigma_t =\sigma_\theta = \sigma_\phi$. It is informative to rewrite the implied total mass profile as
\begin{equation}
\label{eq:massjeans}
M(r) = \frac{r \: \sigma_r^2}{G} \left (\gamma_\star+\gamma_\sigma - 2\beta \right ),
\end{equation}
where $\gamma_{\star} \equiv - \rmd \ln n_\star / \rmd \ln r$ and $\gamma_{\sigma} \equiv - \rmd \ln \sigma_r^2 / \rmd \ln r$. 
Without the benefit of tracer proper motions (or some assumption about the form of the distribution function), the only term on the 
right-hand side of Equation \ref{eq:massjeans} that can be determined by observations is $\gamma_\star$, which follows from 
the projected surface brightness profile under some assumption about how it is related to the projected stellar number density 
$\Sigma_\star(R)$.\footnote{One can make progress if enough individual spectra are obtained such that the population has 
been evenly sampled. However, ensuring that this condition has been met is not trivial.} Via an Abel inversion 
(Equation \ref{eq:abel}) we map $n_\star$ in a one-to-one manner with the spherically deprojected observed surface brightness 
profile (i.e., we assume that the number density traces the light density). As we discuss below, $\sigma_r(r)$ can be inferred from
$\sigmalos(R)$ measurements, but this mapping depends on $\beta(r)$, which is free to vary.

\subsection{Mass Likelihoods from Line-of-Sight Velocity Dispersion Data}
\label{subsec:ML}
Line-of-sight kinematic data provides the projected velocity dispersion profile $\sigmalos(R)$. 
In order to use the Jeans equation one must relate $\sigmalos$ to $\sigma_r$ \citep[as first 
shown by][]{BinneyMamon_82}:
\begin{equation}
\label{eq:LOSrelation}
\Sigma_\star \, \sigmalos^2(R)  =  \int^{\infty}_{R^2} n_\star \sigma_r^2(r) \left[1 - \frac{R^2}{r^2}\beta(r)\right] \frac{\rmd r^2}{\sqrt{r^2 - R^2}}.
\end{equation}
It is clear then that there exists a significant degeneracy associated with using the observed $\Sigma_\star(R)$ and $\sigmalos(R)$ profiles to determine an underlying mass profile $M(r)$ at any radius, as uncertainties in $\beta$ will affect both the mapping between $\sigma_r$ and $\sigmalos$ in Equation \ref{eq:LOSrelation} and the relationship between $M(r)$ and $\sigma_r$  in Equation \ref{eq:massjeans}.  

One technique for handling the $\beta$ degeneracy and providing a fair representation of the allowed mass profile given a set of observables is to consider general parameterizations for $\beta(r)$ and $M(r)$ and then to undertake a maximum likelihood analysis to constrain all possible parameter combinations. In what follows, we use such a strategy to derive meaningful mass likelihoods for a number of dispersion-supported galaxies with line-of-sight velocity data sets. We will use these general results to test our proposed mass estimator. Our general technique is described in the supplementary section of \citet{Strigari_08} and in \citet{Martinez_09}. We refer the reader to these references for a more complete discussion. 

Briefly, for our fiducial procedure we model the stellar velocity dispersion anisotropy as a three-parameter function
\begin{equation}
\beta (r) = (\beta_1 - \beta_0) \frac{r^2}{r^2 + \rbeta^2} + \beta_0, 
\label{eq:betaprofile}
\end{equation}
and model the total mass density distribution using the six-parameter function
\begin{equation} 
\rho_{\rm tot}(r) = \frac{\rho_s \, e^{-r/\rcut}}{(r/r_s)^\gamma [1+(r/r_s)^\alpha]^{(\delta-\gamma)/\alpha}}. 
\label{eq:rhor}
\end{equation}
For our marginalization, we adopt uniform priors over the following
ranges: $\log_{10}(0.2 \, \rhalf) < \log_{10}(\rbeta) < \log_{10}(\rlimit)$;
$-10 < \beta_1 < 0.91$; $-10 < \beta_0 < 0.91$;
$\log_{10}(0.2 \, \rhalf) < \log_{10}(r_s) < \log_{10}(2 \, \rmax)$;
$0 < \gamma < 2$; $3 < \delta < 5$; and $0.5 < \alpha < 3$, where we remind the reader
that $\rlimit$ is the truncation radius for the stellar density.
The variable $\rcut$ allows the dark matter halo profile to truncate at some radius beyond the stellar extent  
and we adopt the uniform prior $\log_{10}(\rlimit) < \log_{10}(\rcut) < \log_{10}(\rmax)$
in our marginalization. For distant galaxies we
use $\rmax = 10 \, \rlimit$ and for satellite galaxies of the Milky Way we
set $\rmax$ equal to the Roche limit for a $10^9 \Msun$ point mass.
In practice, this allowance for $\rcut$ is not important for our purposes because
we focus on integrated masses within the stellar radius.\footnote{We
have explored other prior distributions and find that the results of our
likelihood analysis for $\Mhalf$ are insensitive to these choices.} 
  
%
\begin{figure*}
\includegraphics[width=85mm]{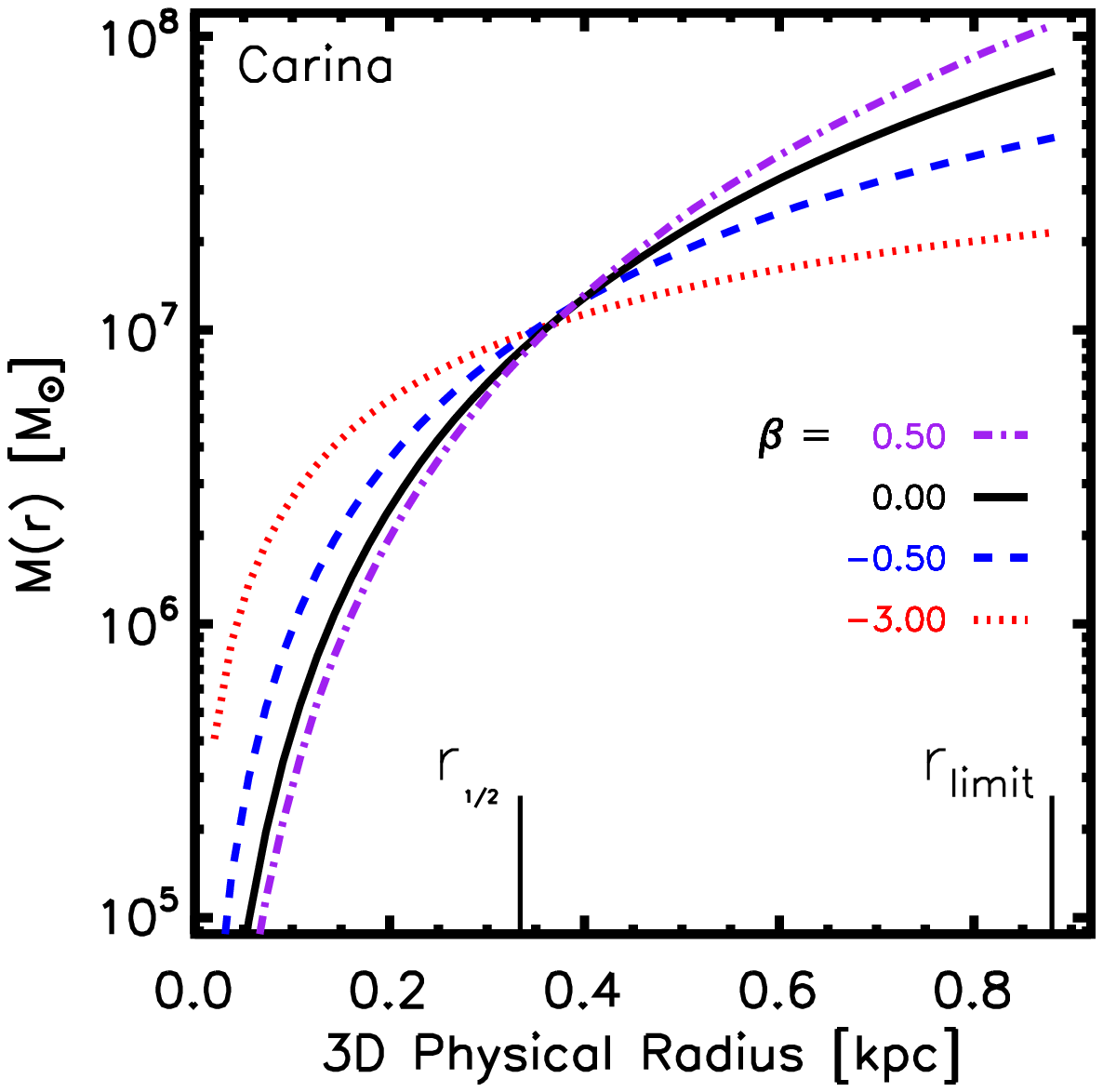}
\hspace{5mm}
\includegraphics[width=85mm]{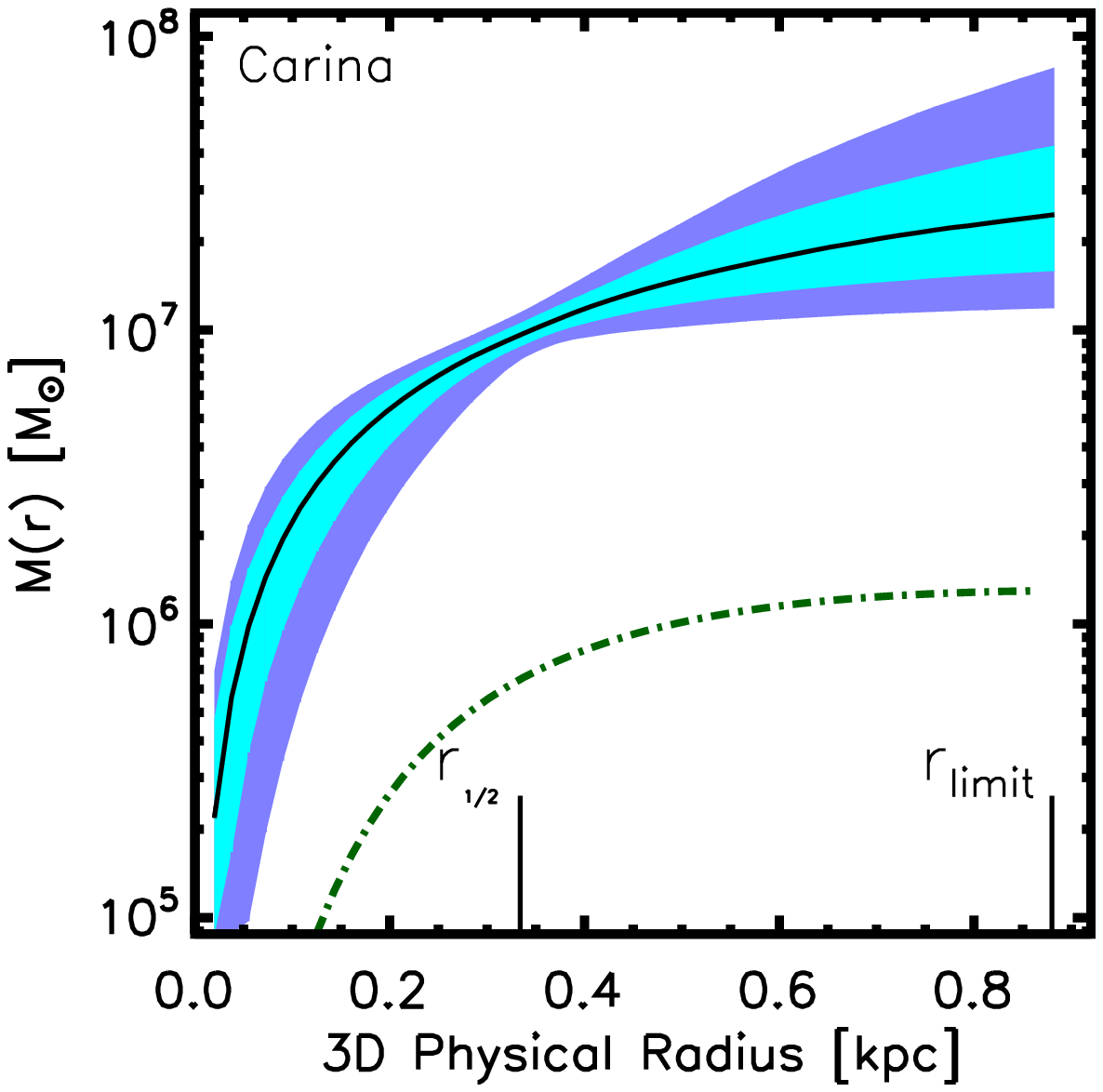}
\caption{{\em Left:} The cumulative mass profile generated by analyzing the Carina dSph using four different constant velocity dispersion anisotropies. The lines represent the median cumulative mass value from the likelihood as a function of physical radius. The width of the mass likelihoods (not shown) do not vary much with radius and are approximately the size of the width at the pinch in the right panel. 
{\em Right:} The cumulative mass profile of the same galaxy, where the black line represents the median mass from our full mass likelihood (which allows for a radially varying anisotropy). The different shades represent the inner two confidence intervals (68\% and 95\%). The green dot-dashed line represents the contribution of mass from the stars, assuming a stellar V-band mass-to-light ratio of 3 $\Msun / \Lsun$.
}
\label{fig:multibeta}
\end{figure*} 
%
  
We also investigate the effects of a more radical model for the stellar velocity
dispersion anisotropy that allows $\beta(r)$ to have an extremum within the
limiting radius. The specific form we use in this second model is 
\beq
\label{eq:betaprofile2}
\beta(r)=\beta_0+(\beta_1-\beta_0)\left(\frac{r}{2\,\rbeta}\right)^2\exp \left[2-\frac{r}{\rbeta}\right],
\eeq
which allows for mild and large variations within the stellar extent depending on the value of 
$\rbeta$. We use the same priors for this functional form as 
those for our fiducial model (Equation \ref{eq:betaprofile}). 
A caveat that bears mentioning is that neither of our $\beta(r)$ profiles 
allow for multiple extrema, but they do allow for large variations
in $\beta(r)$ with radius. Our motivation for investigating these large variations is not based on physical arguments for their existence, but rather to see if the validity of our mass estimator breaks down.

Below we apply our marginalization procedure to resolved kinematic data for MW
dSphs, MW globular clusters, and elliptical galaxies. 
Since MW dSphs and globular clusters are close enough for individual
stars to be resolved, we consider the joint probability of obtaining
each observed stellar velocity given its observational error and the
predicted line-of-sight velocity dispersion from Equations
\ref{eq:jeans} and \ref{eq:LOSrelation}.
In modeling the line-of-sight velocity distribution for any system, we
must take into account that the observed distribution is a convolution
of the intrinsic velocity distribution, arising from the distribution
function, and the measurement uncertainty from each individual star.  
If we assume that the line-of-sight velocity distribution can be
well-described by a Gaussian, which is observationally consistent with the
 best-studied samples
\citep[see, e.g.,][]{Walker_07}, then the probability of obtaining a
set of line-of-sight velocities $\data$ given a set of model
parameters ${\Einastop}$ is described by the likelihood 

\begin{equation}
\label{eq:fulllike}
P(\data| \Einastop) =  \prod_{i=1}^{N} 
\frac{1}{\sqrt{2\pi(\sigma_{\rm th, i}^2 + \epsilon_i^2)}} 
\exp \left[-\frac{1}{2}\frac{(\data_i - \bar{v})^2}{\sigma_{\rm th, i}^2 
+ \epsilon_i^2}\right]\,. 
\end{equation}

The product is over the set of $N$ stars, where $\bar{v}$ is the average velocity of the galaxy. As expected, the total error at a projected position is a sum in quadrature of the theoretical intrinsic dispersion, $\sigma_{\rm th, i}(\Einastop)$, and the measurement error $\epsilon_i$. We generate the posterior probability distribution for the mass at any radius by multiplying the likelihood by the prior distribution for each of the nine  $\beta(r)$ and $\rho_{\rm tot}(r)$ parameters as well as the observationally derived parameters and associated errors that yield $n_\star(r)$ for each galaxy, which include uncertainties in distance. We then integrate over all model parameters, including $\bar{v}$, to derive a likelihood for mass. Following \citet{Martinez_09}, we use a Markov Chain Monte Carlo technique in order to perform the required ten to twelve dimensional integral.\footnote{The volume of parameter space changes depending on the number of free parameters used to fit the photometry of each system, along with the availability of photometric uncertainties. For each MW dSph we have taken care to ensure that we used what we consider to be the most reliable photometry that include observational errors.}
Before moving on, we note that the Gaussian assumption made here is not entirely general, and thus 
is a limiting aspect of our mass modeling.  While most dSph velocity distributions are consistent with Gaussian to within membership errors and errors associated with the possibility of binary star populations \citep{Minor_10}, a small amount of excess kurtosis is measured in the outer parts of some systems \citep{Lokas_09}.

For elliptical galaxies that are located too far for individual stellar spectra to be obtained, we analyze the resolved dispersion profiles with the likelihood
\begin{equation}
\label{eq:dispersionlike}
P(\dataD| \Einastop) = \prod_{i=1}^{N}
\frac{1}{\sqrt{2\pi}\epsilon_i}
\exp \left[-\frac{1}{2}\frac{(\dataD_i - \sigma_{th, i})^2}{\epsilon_i^2}\right]\, ,
\end{equation}
where the product is over the set of $N$ dispersion measurements $\dataD$, and $\epsilon_i$ is the reported error of each measurement. 

\begin{sidewaystable*}
\label{tab:MasterTable}
{\bf Table 1:} Observed and derived properties of spheroidal galaxies considered in this paper.\\
\centering
\begin{tabular}{lccccccccc}
Galaxy & Distance & Luminosity & $\rzero$ & $\rlimit$ & 2D $\Rhalf$ & 3D $\rhalf$ & $\sqrt{\avelos}$ & $\Mhalf$ & $\Upsilon^{\rm V}_{_{1/2}}$\\
& [kpc] & [$\LsunV$] & [arcmin] & [arcmin] & [pc] & [pc] & [$\kms$] & [$\Msun$] & [$\Msun/\LsunV$] \\\hline \hline
Carina (723) & $105 \pm 2\:^{(a)}$ & $4.3^{+1.1}_{-0.9} \times 10^5\:^{(b)}$ & $8.8 \pm 1.2\:^{(c)}$ & $28.8 \pm 3.6\:^{(c)}$ & 254 $\pm$ 28 & 334 $\pm$ 37  & \:\:6.4 $\pm$ 0.2 & $9.56^{+0.95}_{-0.90} \times 10^6$ & 44$^{+13}_{-10}$ \\
Draco  (206) &  $76 \pm 5\:^{(d)}$ & $2.2^{+0.7}_{-0.6} \times 10^5\:^{(b)}$ & $7.63 \pm 0.04\:^{(e)}$ & $45.1 \pm 0.6\:^{(e)}$ & 220 $\pm$ 11 & 291 $\pm$ 14  & 10.1 $\pm$ 0.5 & $2.11^{+0.31}_{-0.31} \times 10^7$ & 200$^{+80}_{-60}$ \\
Fornax  (2409) & $147 \pm 3\:^{(a)}$ & $1.7^{+0.5}_{-0.4} \times 10^7\:^{(b)}$ & $13.7 \pm 1.2\:^{(c)}$ & $71.1 \pm 4.0\:^{(c)}$ & 714 $\pm$ 40 & 944 $\pm$ 53  & 10.7 $\pm$ 0.2 & $7.39^{+0.41}_{-0.36} \times 10^7$ & 8.7$^{+2.8}_{-2.3}$ \\
Leo I  (305) & $254 \pm 18\:^{(f)}$& $5.0^{+1.8}_{-1.3} \times 10^6\:^{(b)}$ & $6.21 \pm 0.95\:^{(g)}$ & $11.70 \pm 0.87\:^{(g)}$ & 295 $\pm$ 49 & 388 $\pm$ 64  & \,\:9.0 $\pm$ 0.4 & $2.21^{+0.24}_{-0.24} \times 10^7$ & 8.8$^{+3.4}_{-2.4}$ \\
Leo II (168) & $233 \pm 15\:^{(h)}$ & $7.8^{+2.5}_{-1.9} \times 10^5\:^{(i)}$ & $2.64 \pm 0.19\:^{(i)}$ & $9.33 \pm 0.47\:^{(i)}$ & 177 $\pm$ 13 & 233 $\pm$ 17  & \,\:6.6 $\pm$ 0.5 & $7.25^{+1.19}_{-1.01} \times 10^6$ & 19$^{+7}_{-5}$ \\
Sculptor (1355) &  $86 \pm 5\:^{(j)}$ & $2.5^{+0.9}_{-0.7} \times 10^6\:^{(b)}$ & $5.8 \pm 1.6\:^{(c)}$ & $76.5 \pm 5.0\:^{(c)}$ & 282 $\pm$ 41 & 375 $\pm$ 54  & \,\:9.0 $\pm$ 0.2 & $2.25^{+0.16}_{-0.15} \times 10^7$ & 18$^{+6}_{-5}$ \\
Sextans  (423) &  $96 \pm 3\:^{(k)}$ & $5.9^{+2.0}_{-1.4} \times 10^5\:^{(b)}$ & $16.6 \pm 1.2\:^{(c)}$ & $160.0 \pm 50.0\:^{(c)}$ & 768 $\pm$ 47 &1019 $\pm$ 62 \:& \,\,7.1 $\pm$ 0.3 & $3.49^{+0.56}_{-0.48} \times 10^7$ & 120$^{+40}_{-35}$ \\
Ursa Minor (212) &  $77 \pm 4 \:^{(l)}$ & $3.9^{+1.7}_{-1.3} \times 10^5\:^{(b)}$ & $17.9 \pm 2.1\:^{(m)}$ & $77.9 \pm 8.9\:^{(m)}$ & 445 $\pm$ 44 & 588 $\pm$ 58  & 11.5 $\pm$ 0.6 & $5.56^{+0.79}_{-0.72} \times 10^7$ & 290$^{+140}_{-90}$ \\ \hline
Bo{\"o}tes I (12) & $66 \pm 3\:^{(n)}$& $2.8^{+0.6}_{-0.4} \times 10^4$ & $7.51^{+0.60}_{-0.54}$& $\sim$ 45 & $242^{+22}_{-20}$ & $322^{+29}_{-27}$ & \:\:9.0 $\pm$ 2.2 & $2.36^{+2.01}_{-1.02} \times 10^7$ & 1700$^{+1400}_{-700}$ \\
Canes Venatici I (214) & $218 \pm 10\:^{(o)}$& $2.3^{+0.4}_{-0.3} \times 10^5$ & $5.30^{+0.24}_{-0.24}$& $\sim$ 50 & $564^{+36}_{-36}$ & $750^{+48}_{-48}$     & \:\:7.6 $\pm$ 0.5 & $2.77^{+0.86}_{-0.62} \times 10^7$ & 240$^{+75}_{-65}$ \\
Canes Venatici II (25) & $160 \pm 5\:^{(p)}$ & $7.9^{+4.4}_{-3.0} \times 10^3$ & $0.95^{+0.18}_{-0.12}$& $\sim$ 10 &\:\;$74^{+14}_{-10}$& \:\;$97^{+18}_{-13}$ & \:\:4.6 $\pm$ 1.0 & $1.43^{+1.01}_{-0.59} \times 10^6$ & 360$^{+380}_{-180}$ \\
Coma Berenices (59) &  $44 \pm 4\:^{(q)}$ & $3.7^{+2.2}_{-1.4} \times 10^3$ & $3.57^{+0.36}_{-0.36}$& $\sim$ 18 &\:\;$77^{+10}_{-10}$& $100^{+13}_{-13}$    & \:\:4.6 $\pm$ 0.8 & $1.97^{+0.88}_{-0.60} \times 10^6$ & 1100$^{+800}_{-500}$ \\
Hercules $^{(r)}$ (30) & $133 \pm 6$& $1.1^{+0.5}_{-0.3} \times 10^4$ & $3.52^{+0.30}_{-0.30}$& $\sim$ 40 & $229^{+19}_{-19}$  & $305^{+26}_{-26}$    & \:\:5.1 $\pm$ 0.9 & $7.50^{+5.72}_{-3.14} \times 10^6$ & 1400$^{+1200}_{-700}$ \\
Leo IV (17) & $160 \pm 15\:^{(q)}$& $8.7^{+5.4}_{-3.6} \times 10^3$ & $1.49^{+0.30}_{-0.42}$& $\sim$ 15 & $116^{+26}_{-34}$  & $151^{+34}_{-44}$    & \:\:3.3 $\pm$ 1.7 & $1.14^{+3.50}_{-0.92} \times 10^6$ & 260$^{+1000}_{-200}$ \\
Leo T $^{(s)}$ (18) & $407 \pm 38$& $1.4 \times 10^5$ & $0.68^{+0.08}_{-0.08}$ & $4.8 \pm 1.0$ & $115^{+17}_{-17}$  & $152^{+21}_{-21}$ & \:\:7.8 $\pm$ 1.6 & $7.37^{+4.84}_{-2.96} \times 10^6$ & 110$^{+70}_{-40}$ \\
Segue 1 (24) &  $23 \pm 2\:^{(q)}$ & $3.4^{+3.0}_{-1.6} \times 10^2$ & $2.62^{+0.71}_{-0.36}$& $\sim$ 20 & \,$29^{+8}_{-5}$   & \:\;$38^{+10}_{-7}$  & \:\:4.3 $\pm$ 1.1 & $6.01^{+5.07}_{-2.80} \times 10^5$ & 3500$^{+5000}_{-2000}$ \\
Ursa Major I (39) &  $97 \pm 4\:^{(t)}$ & $1.4^{+0.4}_{-0.4} \times 10^4$ & $6.73^{+1.01}_{-0.77}$& $\sim$ 50 & $318^{+50}_{-39}$  & $416^{+65}_{-51}$    & \:\:7.6 $\pm$ 1.0 & $1.26^{+0.76}_{-0.43} \times 10^7$ & 1800$^{+1300}_{-700}$ \\
Ursa Major II (20) &  $32 \pm 4\:^{(u)}$ & $4.0^{+2.5}_{-1.4} \times 10^3$ & $9.52^{+0.60}_{-0.60}$& $\sim$ 50 & $140^{+25}_{-25}$  & $184^{+33}_{-33}$    & \:\:6.7 $\pm$ 1.4 & $7.91^{+5.59}_{-3.14} \times 10^6$ & 4000$^{+3700}_{-2100}$ \\
Willman 1 (40) &  $38 \pm 7\:^{(v)}$ & $1.0^{+0.9}_{-0.5} \times 10^3$ & $1.37^{+0.12}_{-0.24}$& $\sim$ 9 & \,$25^{+5}_{-6}$    & $33^{+7}_{-8}$       & \:\:4.0 $\pm$ 0.9 & $3.86^{+2.49}_{-1.60} \times 10^5$ & 770$^{+930}_{-440}$ \\
\hline
NGC 185 (n=1.2$\:^{(w)}$) & $616 \pm 26\:^{(x)}$ & $1.1 \times 10^8\:^{(y*)}$ & $1.49\:^{(y)}$ & $\sim 14.9$ & 266 & 355 & 31 $\pm$ 1 & $2.93^{+1.02}_{-0.77} \times 10^8$ & $5.3^{+1.9}_{-1.4}$ \\
NGC 855 (n=1.9$\:^{(w)}$) & $9320\:^{(z)}$ & $1.1 \times 10^9\:^{(aa*)}$ & $0.23\:^{(aa)}$ & $\sim 2.30$ & 624 & 837 & 58 $\pm$ 3 & $2.48^{+0.54}_{-0.49} \times 10^9$&$4.5^{+1.0}_{-0.9}$\\
\hline
NGC 499 (n=3.6$\:^{(w)}$) & $62300\:^{(z)}$ & $4.1 \times 10^{10}\:^{(bb*)}$ & $0.25\:^{(bb)}$ & $\sim 2.50$ & 4500 & 6070 & 274 $\pm$ 7 & $3.27^{+0.48}_{-0.54} \times 10^{11}$ & $16^{+2.3}_{-2.6}$ \\
NGC 731 (n=3.8$\:^{(w)}$) & $52700\:^{(z)}$ & $3.9 \times 10^{10}\:^{(aa*)}$ & $0.24\:^{(aa)}$ & $\sim 2.40$ & 3600 & 4850 & 163 $\pm$ 1 & $8.52^{+1.06}_{-0.89} \times 10^{10}$ & $4.4^{+0.5}_{-0.5}$ \\
NGC 3853 (n=4.0$\:^{(w)}$) & $44600\:^{(z)}$ & $2.1 \times 10^{10}\:^{(cc*)}$ & $0.24\:^{(cc)}$ & $\sim 2.40$ & 3050 & 4110 & 198 $\pm$ 3 & $8.54^{+1.28}_{-1.49} \times 10^{10}$ & $8.1^{+1.2}_{-1.4}$ \\
NGC 4478 (n=2.07$\:^{(dd)}$) & $16980\:^{(dd)}$ & $7.0 \times 10^9\:^{(dd)}$ & $0.22\:^{(dd)}$ & $1.73\:^{(dd)}$ & 1110 & 1490 & 147 $\pm$ 1 & $1.96^{+0.23}_{-0.28} \times 10^{10}$ & $5.6^{+0.7}_{-0.8}$ \\
\end{tabular}
\vskip 0.5 cm Note: Galaxies are grouped from top to bottom as pre-SDSS/classical MW dSphs, post-SDSS MW dSphs, dwarf elliptical galaxies (dEs), and elliptical galaxies (Es). Within the parentheses next to each MW dSph is the number of stars analyzed. The dSphs with errors on $\rlimit$ are fit with King profiles (where $\rzero = \rcore$). Those without sources for $\rlimit$ are estimated from Figure 1 of \citet{Martin_08b} (we found that our $\Mhalf$ determinations were largely insensitive to the choice of reasonable $\rlimit$ values). Except for Leo T, all of the post-SDSS dwarfs are fit with truncated exponential light distributions (where $\rzero$ is the exponential scale length derived from the half-light radius). The dEs and Es are fit with truncated S\'ersic profiles, where each limiting radius is not usually quoted in the literature. Also note that errors on the masses are approximately normal in log$_{10}$($\Mhalf$). Lastly, note that the quoted errors in the luminosities and in the dynamical mass-to-light ratios were derived in this paper and are also approximately log-normal. For the classical dSphs we took into account the errors in the apparent magnitudes and the errors in the distance estimates. For the post-SDSS dSphs we considered the quoted errors in absolute magnitudes.\\
References: Values in column 5 (2D $\Rhalf$) for the classical MW dSphs and Leo T, and the values in columns 6-9 for all of the MW dSphs are derived in this paper from the quoted elliptical fits to the surface brightness profiles from the cited sources (this convention differs from the geometric means that are sometimes quoted from the equivalent elliptical fits \citep[see, e.g., Section 3 of][]{Irwin_95}. Except for Hercules and Leo T, values in columns 2-5 of the post-SDSS MW dSphs are from \citet{Martin_08b}. Lastly, the values in columns 5-9 for the dEs and Es are derived in this paper.
The individual references are as follows: a) \citet{Pietrzynski_09} b)
Rederived from apparent magnitudes listed in \citet{Mateo_98}, c)
\citet{Irwin_95}, d) \citet{Bonanos_04}, e) \citet{Segall_07}, f)
\citet{Bellazzini_04}, g) \citet{Smolcic_07}, h)
\citet{Bellazzini_05}, i) \citet{Coleman_07}, j)
\citet{Pietrzynski_08}, k) \citet{Lee_03}, l) \citet{Carrera_02}, m)
RGB tracers from \citet{Palma_03}, n) \citet{DallOra_06}, o) \citet{Martin_08a}, p) \citet{Greco_08}, q) \citet{belokurov07}, r) \citet{Sand_09}, s) \citet{deJong_08}, t) \citet{Okamoto_08}, u) \citet{zucker06a}, v) \citet{willman05a}, w) Derived from \citet{PandH_98}, x) \citet{McConnachie_05}, y) \citet{SandP_02}, z) Quoted from NASA/IPAC Extragalactic Database, aa) \citet{SandP_00}, bb) \citet{SandP_97c}, cc) \citet{SandP_97b}, dd) \citet{Kormendy_09}, who present similar parameters to those the originally derived in \citet{Ferrarese_06}. *)Luminosities derived from applying $B-V$ values calculated in \citet{Fukugita_95}. Lastly, the references for the kinematic data used to derive the velocity dispersions are listed in the caption of Figure \ref{fig:Mrhalf}.
\end{sidewaystable*}

\section{Minimizing the Anisotropy Degeneracy}
\label{sec:mass}
\subsection{Expectations}

Qualitatively, one might expect that the degeneracy between the integrated mass and the assumed anisotropy parameter will be minimized
at some intermediate radius within the stellar distribution. Such an expectation follows from considering the relationship between $\sigmalos$ and $\sigma_r$.  

At the projected center of a spherical, dispersion-supported galaxy
($R = 0$), line-of-sight observations project onto the radial
component with $\sigmalos \sim \sigma_r$, while at the edge of the
galaxy ($R = \rlimit$), line-of-sight velocities project onto the
tangential component with $\sigmalos \sim \sigma_t$. For example,
consider a galaxy that is intrinsically isotropic ($\beta = 0$). If
this system is analyzed using line-of-sight velocities under the false
assumption that $\sigma_r > \sigma_t$ ($\beta > 0$) at all radii, then 
the total velocity dispersion at $r \simeq 0$ would be underestimated 
while the total velocity dispersion at $r \simeq \rlimit$ would be
overestimated. Conversely, if one were to analyze the same galaxy
under the assumption that $\sigma_r < \sigma_t$ ($\beta < 0$) at all radii,
then the total velocity dispersion would be overestimated near the center
and underestimated near the galaxy edge. It is plausible then that
there exists some intermediate radius where attempting to infer the
enclosed mass from only line-of-sight kinematics is minimally affected
by the unknown value of $\beta$. 

These qualitative expectations are borne out explicitly in Figure
\ref{fig:multibeta}, where we present inferred mass profiles for the
Carina dSph galaxy for several choices of constant $\beta$. The
right-hand panel shows the same data analyzed using our full
likelihood analysis, where we marginalize over the fairly general
$\beta(r)$ profile presented in Equation \ref{eq:betaprofile}.
We use 723 stellar velocities from \citet{Walker_09a} with the
constraint that their membership probabilities (which are based 
on a combination of stellar velocity and metallicity)
are greater than 0.9, and in projection they lie within
650 pc of the center (which is below the lower limit of $\rlimit$
given in Table 1). The average velocity error of this set is
approximately 3 $\kms$. Each line in the left panel of Figure \ref{fig:multibeta} shows
the median likelihood of the cumulative mass value at each radius for
the value of $\beta$ indicated. The 3D half-light radius and the limiting
stellar radius are marked for reference. As expected, forcing $\beta > 0$
produces a systematically lower (higher) mass at a small (large)
radius compared to $\beta < 0$. This of course demands that every pair
of $M(r)$ profiles analyzed with different assumptions about $\beta$ cross 
at some intermediate radius.\footnote{\citet{VDM_00}
demonstrated a comparable result with more restrictive conditions.}
Somewhat remarkable is the fact that every pair intersects at approximately the same
radius. We see that this radius is very close to the deprojected 3D
half-light radius $\rhalf$. The right-hand panel in Figure
\ref{fig:multibeta} shows the full mass likelihood as a function of
radius (which allows for a radially varying anisotropy), where the shaded bands illustrate the 68\% and 95\% likelihood
contours, respectively. The likelihood contour also pinches near
$\rhalf$, as this mass value is the most constrained by the data. 

By examining each of the well-sampled dSph kinematic data sets \citep{Munoz_05, Koch_07, Mateo_08, Walker_09a}
in more detail, we find that the error on mass near $\rhalf$ is always dominated by measurement errors
(including the finite number of stars) rather than the $\beta$
uncertainty, while the mass errors at {\em both} smaller and larger radii are
dominated by the $\beta$ uncertainty (and thus are less affected by
measurement error).\footnote{A similar effect was discussed but not 
fully explored in \citet{Strigari_07b}.} We now explain this result
by examining the Jeans equation in the context of observables. 

\subsection{Why is the mass within half-light radius insensitive to
velocity dispersion anisotropy?}
\label{subsec:magicradius}

Here we present the derivation of Equations \ref{eq:main2} and \ref{eq:main}. We start by
analytically showing that there exists a radius $\req$ within which the
dynamical mass will be minimally affected by the velocity dispersion anisotropy,
$\beta(r)$. We then consider two cases of interest for observed
dispersion-supported systems. First, we consider the case when the  
velocity dispersion anisotropy is spatially constant and show that
$\req \simeq \rthree$ where $\rthree$ is an observable defined such that
$\gamma_\star \equiv -{\rm d} \ln n_\star / {\rm d} \ln r = 3$ at $r=\rthree$. Second, we extend
our analysis to allow for non-constant $\beta(r)$ and show that under
mild assumptions about the variation of $\beta(r)$, the mass within
radius $\rthree$ is insensitive to the velocity dispersion
anisotropy.

While the steps outlined above provide a deeper insight into Equation
\ref{eq:main2}, the essence of our arguments can be laid out in a
few lines. We begin by rewriting the Jeans equation such that
the $\beta(r)$ dependence is absorbed into the definition of 
$\sigmatot^2 = \sigma_r^2 + \sigma_\theta^2 + \sigma_\phi^2 = (3 - 2 \beta)\sigma_r^2$:
\begin{equation}
\label{eq:massjeans2}
G M(r) r^{-1} = \sigmatot^2(r) +  \sigma_r^2(r) \left (\gamma_\star+\gamma_\sigma -3 \right).
\end{equation}
We then note that if $\gamma_\sigma(\rthree) \ll 3$ (as our numerical
computations show it must be for flat observed $\sigmalos(R)$
profiles), then at $r=\rthree$ the mass depends only on $\sigmatot$ and
we may write 
\begin{eqnarray}
\label{eq:mr3}
M(\rthree) & \simeq & G^{-1}\sigmatot^2(\rthree) \, \rthree \simeq G^{-1} \ave{\sigmatot^2} \, \rthree \\ \nonumber
& \simeq & 3 \, G^{-1} \avelos \, \rthree \,,
\end{eqnarray}
where the last line is Equation \ref{eq:main2}. We remind the reader that the brackets indicate 
a luminosity-weighted average over the entire system. In the above chain of
arguments we have used the relation $\ave{\sigmatot^2}\simeq\sigmatot^2(\rthree)$.
We will show why this is a good approximation in Section \ref{sec:constantbeta}.  
 
Finally, we show in Appendix \ref{ap:conversions} that the log-slope of $n_\star$ 
is approximately 3 at the deprojected half-light radius $\rthree \simeq \rhalf$ 
for most common light profiles, and therefore the last line of Equation \ref{eq:mr3}
provides our mass estimator (Equation \ref{eq:main}). For example, $\rthree
\simeq 0.94 \, \rhalf$ for a Plummer profile and $\rthree \simeq
1.15 \, \rhalf$ for \citet{King_62} profiles and for the family of
\citet{Sersic_68} profiles with $n=0.5$ to $10$. The relationships between $\rhalf$ and the
observable scale radii for various commonly-used surface density
profiles are provided in Appendix \ref{ap:conversions}.

\subsubsection{Existence of a radius $\req$ where the mass profile 
is minimally affected by anisotropy}

Consider a velocity dispersion-supported stellar system that is well studied,
such that $\Sigma_\star(R)$ and $\sigmalos(R)$ are determined accurately by
observations.  If we model this system's mass profile using the Jeans
equation, any viable solution will keep the quantity $\Sigma_\star(R) \,
\sigmalos^2(R)$ fixed to within allowable errors. With this in mind,  
we rewrite Equation \ref{eq:LOSrelation} in a form that is invertible,
isolating the integral's $R$-dependence into a kernel: 
\beq
\label{eq:LOSstep}
\Sigma_\star \sigmalos^2(R)  =  \int_{R^2}^{\infty} \left[\frac{n_\star \sigma_r^2}
{(1 - \beta)^{-1}} + \int_{r^2}^\infty \frac{\beta n_\star \sigma_r^2}{2\tilde{r}^2} 
\rmd \tilde{r}^2 \right] \frac{\rmd r^2}{\sqrt{r^2-R^2}}.
\eeq
We explain this derivation in Appendix A, where we also perform an
Abel inversion to solve for $\sigma_r(r)$ and $M(r)$ in terms of
directly observable quantities (while we were writing this paper we learned
that Mamon \& Bou{\'e} 2010 had independently performed a similar analysis.)

Because Equation \ref{eq:LOSstep} is invertible, the fact that the
left-hand side is an observed quantity and independent of $\beta$
implies that the term in brackets must be well determined regardless 
of a chosen $\beta$. This allows us to equate the isotropic integrand
with an arbitrary anisotropic integrand: 
\beq
\label{eq:integrand}
 \left . n_\star \sigma_r^2 \right |_{\beta = 0} = n_\star
 \sigma_r^2 [1-\beta(r)] + \int_r^\infty \frac{\beta n_\star
   \sigma_r^2 \rmd \tilde{r}}{\tilde{r}}. 
\eeq
We now take a derivative with respect to $\ln r$ and subtract Equation 
\ref{eq:jeans}
to obtain the following result
\begin{eqnarray}
\label{eq:massdiff}
M(r; \beta) - M(r; 0) = \frac{\beta(r) \: r \: \sigma_r^2(r)}{G} 
\left( \gamma_\star + \gamma_\sigma + \gamma_\beta - 3 \right).
\end{eqnarray}
We remind the reader that $\gamma_\star \equiv - \rmd \ln n_\star / \rmd \ln r$ 
and $\gamma_\sigma \equiv - \rmd \ln \sigma_r^2 / \rmd \ln r$. Following the same 
nomenclature, $\gamma_\beta \equiv - \rmd \ln \beta / \rmd \ln r = - \beta^\prime/\beta$, 
where $^\prime$ denotes a derivative with respect to $\ln r$.

Equation \ref{eq:massdiff} reveals the possibility of a radius $\req$ where the term in parentheses 
goes to zero, such that the enclosed mass $M(\req)$ is minimally affected by our ignorance of 
$\beta(r)$~\footnote{For $\beta$ profiles that are close to isotropic, solving for $\req$ is not 
necessary, as the right-hand side of Equation \ref{eq:massdiff} is close to 0 everywhere.}:
\begin{equation}
\label{eq:req}
\gamma_\star(\req) = 3 - \gamma_\sigma(\req) - \gamma_\beta(\req) \,.
\end{equation}
While in principle one needs to know $\gamma_\beta$ in order to determine $\req$, 
we argue below that this term must be small for realistic cases that correspond to 
observed galaxies.\footnote{Note that for anisotropic parameterizations that become 
close to isotropic, $\gamma_\beta$ may be large. However, the combination 
$\beta \: \gamma_\beta = \beta^\prime$ is still well-behaved.}
Given this, a solution for $\req$ must exist. One can see this immediately,
as analyzing the luminosity-weighted\footnote{The integral is actually 
number-weighted, but we map number density to luminosity density in a one-to-one manner.} 
average of Equation \ref{eq:massjeans2} in conjunction with the scalar virial 
theorem (Equation \ref{eq:vt}) requires that 
$\langle (\gamma_\star + \gamma_\sigma -3) \sigma_r^2 \rangle = 0$.
Since $\sigma_r^2(r)$ is positive definite, it must be true that there exists
at least one radius where $\gamma_\star = 3 - \gamma_\sigma$.
More specifically, for typically observed stellar profiles, $\gamma_\star(r)$
changes from being close to zero (cored) in the center
to larger than 3 in the outer parts (to keep the stellar mass
finite). (For example, $\gamma_\star$ for a Plummer profile
transitions from 0 to 5.) The changes in $\gamma_\sigma(r)$ are more
benign (see Equation \ref{eq:rhostar}). Putting these facts together, we see 
that unless $\gamma_\beta$ is very large in magnitude, Equation
\ref{eq:req} will have a solution.

In order to determine the value of $M(\req)$ we manipulate 
Equation \ref{eq:sigmaprime2} in order to isolate the relationship
between $\sigma_r^2(r)$ and $\avelos$.
\beq
\label{eq:sigmaprime}
\gammao(r)\avelos = \left[(\gamma_\star + \gamma_\sigma)(1-\beta) + \beta + \beta^\prime\right]\sigma_r^2.
\eeq
Here, the quantity $\gammao(r)$ is dimensionless and depends only observable functions:
\beq
\gammao(r) \equiv 
\frac{1}{n_\star(r)\avelos\pi}\left(\int_{r^2}^\infty
\frac{\rmd (\Sigma_\star \sigmalos^2)}{\rmd R^2}
\frac{\rmd R^2}{\sqrt{R^2 - r^2}} \right)^\prime \,.
\eeq
Note that in the limit where $\sigmalos$ is constant we have $\gammao(r)=\gamma_\star(r)$, 
which arises by utilizing an Abel inversion (Equation \ref{eq:abel}).
Now we may use Equations \ref{eq:massjeans2}, \ref{eq:req}, and \ref{eq:sigmaprime} to show
\begin{equation}
\label{eq:Meq}
M(\req) = \gammao(\req) \, G^{-1} \, \avelos \, \req \,.
\end{equation}

As mentioned above, for generic cases the value of
$\req$ will depend on $\beta(r)$ and thus
our ignorance of $\beta(r)$ is now translated to $\req$. However, as we discuss in the
next section, if the observed 
$\sigmalos(R)$ does not vary much compared to $\Sigma_\star(R)$ 
(as is true for most spheroidal systems), then $\req \simeq \rthree$ 
and $\gammao(\req)\simeq 3$. More generally, each galaxy will have a
  different $\req$, which can be searched for numerically using
  Equation \ref{eq:massdiff} in conjunction with the family of $M(r)$
  and $\beta(r)$ profiles that solve the Jeans equation. 
When we actually  perform this analysis on real galaxies  using our maximum likelihood
  approach, we find that the likelihoods for $\req$ peak near
  $\rthree \simeq \rhalf$. 
  
\subsubsection{Spatially constant velocity dispersion anisotropy\label{sec:constantbeta}}

In this section, we assume that $\beta(r)$ is constant and show that 
$\req$ is close to $\rthree$.  
We start with Equation \ref{eq:sigmaprime} and set $\beta^\prime=0$ to
yield:  
\begin{equation}
\label{eq:gammasig}
\gamma_\sigma(\rthree)\sigmatot^2(\rthree)\frac{1-\beta}{3-2\beta} \simeq
3\avelos - \sigmatot^2(\rthree)\,.
\end{equation}
We have assumed that $\sigmalos$ varies slowly with radius
such that $\gammao \simeq 3$.  Of course, physically, $\sigmalos$ has to decrease 
as $R$ approaches the stellar limiting radius, but we find numerically that the
relation above is still a good approximation as long as the variations
in the observed $\sigmalos$ are mild at $R \simeq \Reff$. Equation \ref{eq:gammasig} 
tells us that if $\gamma_\sigma(\rthree)$ is small and $\beta$ is constant, then
$\sigmatot^2(\rthree)\simeq 3\avelos$. This provides one justification  
for the second step in Equation \ref{eq:mr3}.

We now turn to a more detailed computation of $\sigmatot^2(\rthree)$ to 
elucidate the role of $\gamma_\sigma$, without explicitly assuming
that $\sigmalos(R)$ is constant. Consider the average total
velocity dispersion written explicitly as an integral over
$\sigma_r^2$,   
\begin{equation}
\label{eq:sigtot}
\ave{\sigmatot^2} \, =  \, 4 \pi \, \int_{-\infty}^\infty r^3 \, n_\star \,
\sigma_r^2 \, (3-2\beta) \, \rmd \ln r.
\end{equation}

In realistic cases, $n_\star$ will vary significantly with radius from a
flat inner profile with $\gamma_\star = 0$  at small $r$ to
a steep profile with $\gamma_\star > 3$ at large r. Thus the integrand
is expected to be single peaked unless $\sigma_r$ varies
in an unexpectedly strong way to compensate for the behavior of $n_\star$.
However, since observed $\sigmalos$ profiles do not vary much
with position in the sky, $\sigma_r(r)$ must also vary smoothly
with radius (at least for constant $\beta$; see Equation \ref{eq:rhostaralt}). 
Thus the integrand will peak at $r=\rsig$ such that
$\gamma_\star(\rsig)+\gamma_\sigma(\rsig)=3$.  
We may then  use a saddle point approximation after a Taylor expansion of 
the natural logarithm of the integrand about $\rsig$, approximating
the integral as a Gaussian
\begin{eqnarray}
\ave{\sigmatot^2}  &\simeq &  4 \pi A(\rsig)
\int_{-\infty}^\infty \exp \left[-\frac{K(\rsig)}{2}
\left(\ln \left[\frac{r}{\rsig}\right] \right)^2  \right] \rmd \ln r \nonumber\\
& \simeq & 4 \pi \sqrt{\frac{2 \pi}{K(\rsig)}}
A(r_\sigma)\,.
\label{eq:forma}
\end{eqnarray}
where
\begin{eqnarray}
\label{eq:defofA}
A(r) \, = \, r^3 \, n_\star(r) \: \sigmatot^2(r), \;{\rm and}\;
K(r)=\gamma_\star^\prime(r)+\gamma_\sigma^\prime(r)\,. 
\end{eqnarray}

Similarly, since $r^3 \, n_\star$ peaks at $\rthree$, one can repeat the analysis of the previous paragraph to write
\begin{equation}
\label{eq:normalizeN}
1 = 4 \pi \int_{-\infty}^\infty r^3 n_\star \, \rmd \ln r \simeq 4 \pi
\sqrt{\frac{2 \pi}{\gamma_\star^\prime(\rthree)}} \rthree^3 n_\star(\rthree) .
\end{equation}
The term $A(\rsig)$ computed at $\gamma_\star+\gamma_\sigma=3$ is 
different from $A(\rthree)$ at second order in $\gamma_\sigma(\rthree)$. Thus,
even for moderate values of $\gamma_\sigma(\rthree)$ we may replace
$A(\rsig)$ in Equation \ref{eq:forma} with $A(\rthree)$ to find (with the
aid of Equations \ref{eq:defofA} and \ref{eq:normalizeN}):
\begin{equation}
\label{eq:avelos2}
3\avelos=\ave{\sigmatot^2} \, \simeq \,
\sqrt{\frac{\gamma_\star^\prime(\rthree)}{\gamma_\star^\prime(\rsig) +
\gamma_\sigma^\prime(\rsig)}} \sigmatot^2(\rthree) \, \simeq \,
\sigmatot^2(\rthree). 
\end{equation}
The last approximation arises by neglecting the first order correction in
$\gamma_\sigma$, enabling us to evaluate the terms inside of the square root
at $r=\rthree$. Our numerical mass estimates show that the observational error
is larger than that due to the neglect of the $\gamma_\sigma$
term.

Next we take the derivative of Equation \ref{eq:sigmaprime} at $r=\rthree$:
\beq
\label{eq:sigmaprime3}
\gamma_\star^\prime(\rthree)+\gamma_\sigma^\prime(\rthree) \simeq \gamma_\star^\prime(\rthree)
\, \frac{3-2\beta}{3-3\beta} \: ,
\eeq
where we have neglected $\gamma_\sigma(\rthree)$. From this expression, 
we see that it is only for values of $\beta$ close to unity that the last step in
Equation \ref{eq:avelos2} is not a good approximation. Such large
values of constant $\beta$, however, are disfavored by the Jeans equation when considering
realistic dispersion profiles. This may be seen by taking a derivative of the Jeans
equation (Equation \ref{eq:massjeans2}) at $r=\rthree$ to write
\beq
\gamma_\star^\prime(\rthree)+\gamma_\sigma^\prime(\rthree) 
\simeq (3-2\beta)(2-\gamma_\rho),
\eeq
where we neglected the $\gamma_\sigma(\rthree)$ term and where we set 
$M(r) = M(\rthree) (r/\rthree)^{3-\gamma_\rho}$. Combining this with 
Equation \ref{eq:sigmaprime3}, we require that
\beq
\label{eq:betadisfav}
1-\beta \simeq \frac{\gamma_\star^\prime(\rthree)}{6 - 3 \, \gamma_\rho},
\eeq
which shows that $\beta$ values close to 1 are disfavored because
observations reveal that $\gamma_\star^\prime(\rthree)$ is of order unity
for systems in equilibrium.\footnote{Note that if 
$\gamma_\rho > 2$, Equation \ref{eq:betadisfav} 
yields the unphysical result of $\beta > 1$, implying that 
$\gamma_\sigma(\rthree)$ should not be neglected.} 
With regard to large negative $\beta$ values,
these extremes are preferred when $\gamma_\rho \lesssim 2$. We remind 
the reader that in the above arguments we have neglected
$\gamma_\sigma(\rthree)$ in keeping with our focus on systems with flat observed
velocity dispersion profiles (see Equation \ref{eq:rhostaralt}).


As an aside, we note that even if we knew $\beta(r)$, uncertainties in the 
inner stellar profile will limit how well we recover the slope of the total 
density profile $\gamma_\rho$ at $\rthree$. 

Given this, Equation \ref{eq:avelos2} can be considered a good approximation. 
That is, $3\avelos \simeq \sigmatot^2(\rthree)$ if $\beta$ is constant and as
long as the observed $\sigmalos$ does not vary much with 
position on the sky. Our full numerical analysis of observed
spectroscopic data show that this is indeed the preferred solution 
of the Jeans equation. This realization, together with Equation \ref{eq:massjeans2},
allows us to derive our mass estimator presented in Equation \ref{eq:main}, 
with $\rhalf \simeq \rthree$. 
%
\begin{figure*}
\includegraphics[width=85mm]{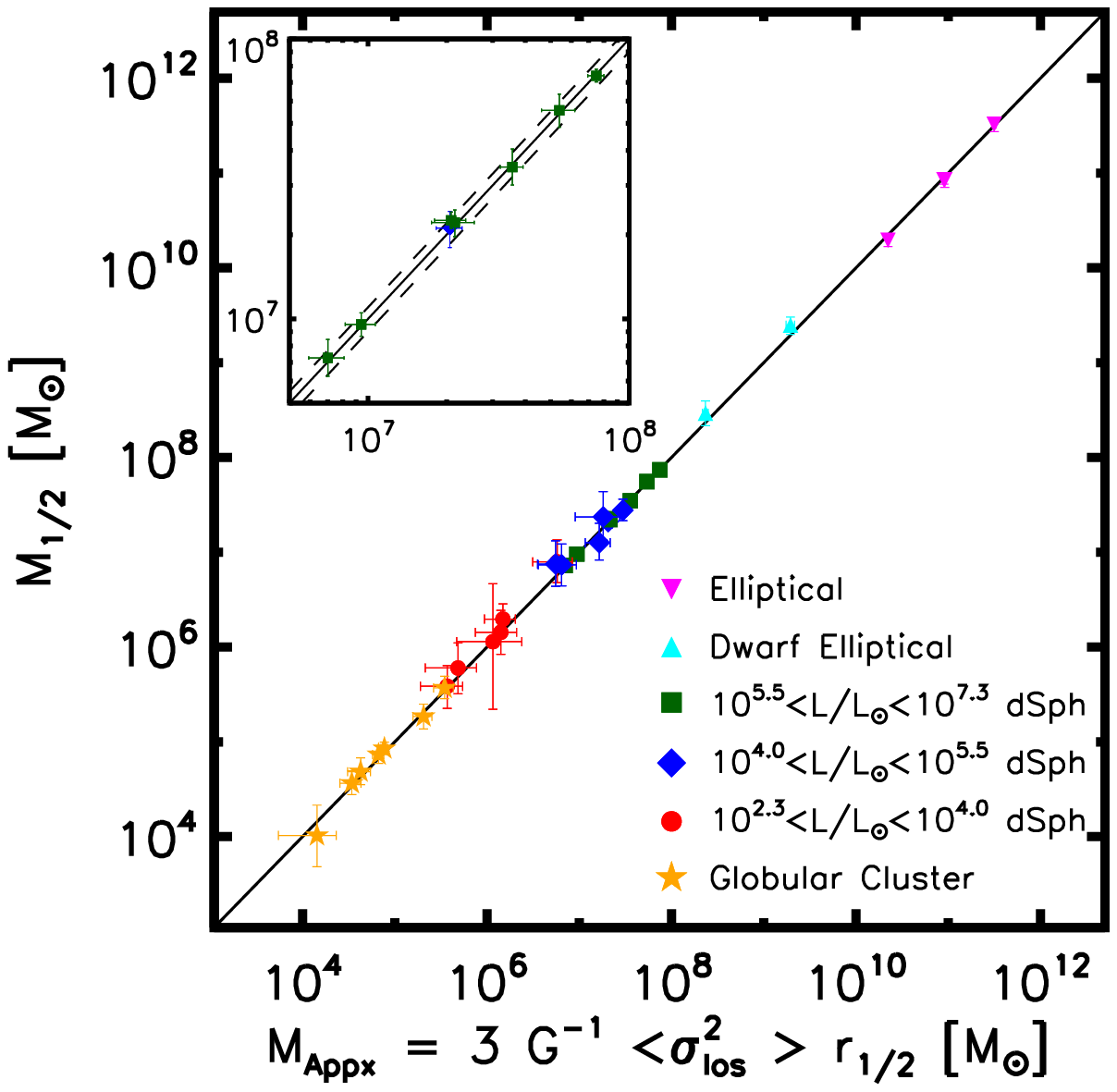}
\hspace{5mm}
\includegraphics[width=85mm]{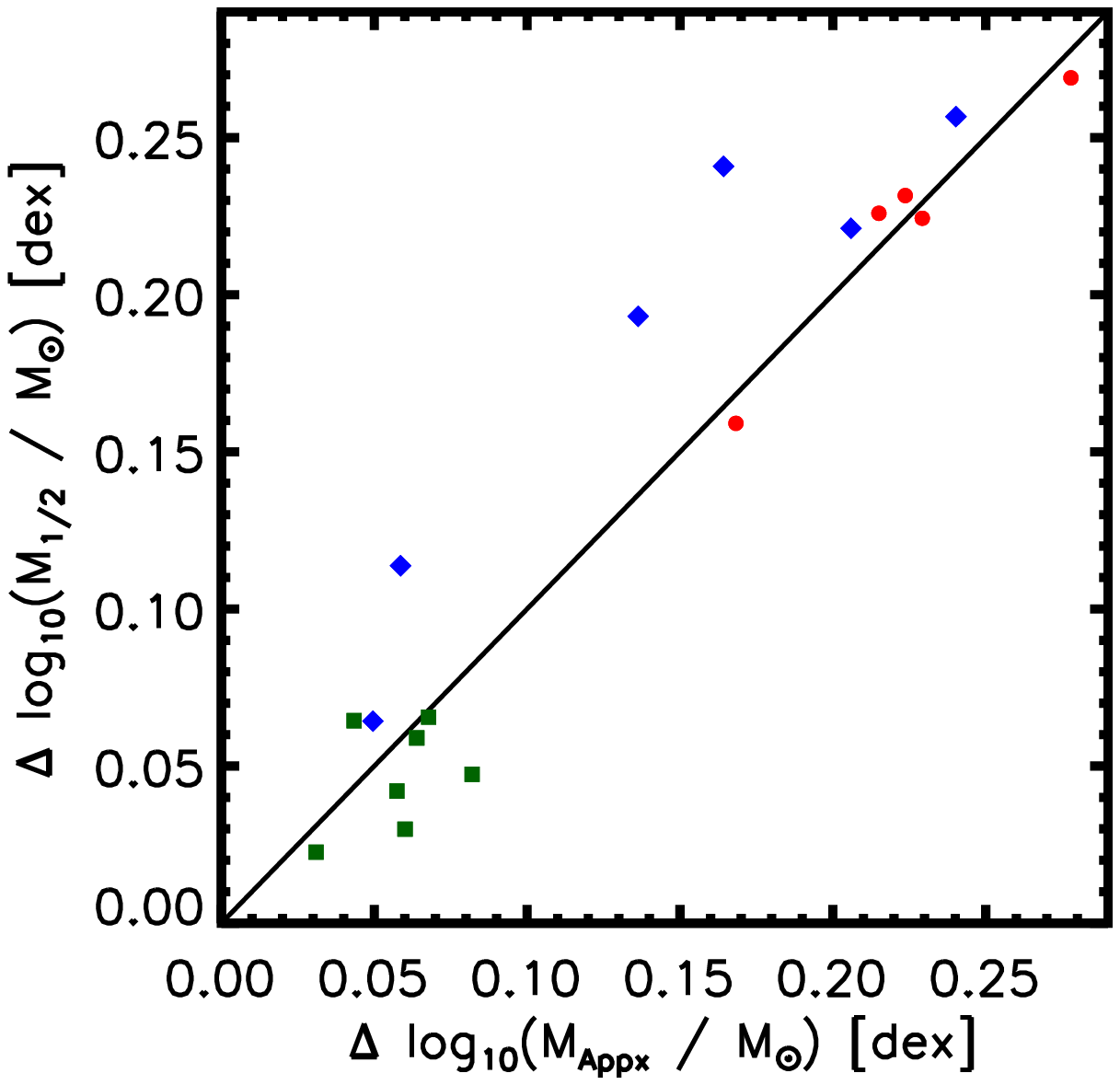}
\caption{
{\em Left:} The half-light masses for Milky Way dSphs (green squares, blue diamonds, red circles), galactic globular clusters (yellow stars), dwarf ellipticals (cyan triangles), and ellipticals (pink inverted triangles). The vertical axis shows masses obtained using our full likelihood analysis. The horizontal axis shows mass estimates based on our mass estimator, Equation \ref{eq:main}. The inset focuses on the pre-SDSS (classical) dSphs, where the dotted lines indicate a 10\% scatter in our mass estimator.
{\em Right:} Errors on half-light masses for Milky Way dSphs. The vertical axis shows the 68\% error width derived from our full likelihood analysis and the horizontal axis shows the error width calculated by straightforward error propagation using Equation \ref{eq:main}. The agreement between the two demonstrates that errors on the mass determinations within the 3D deprojected half-light radius $\rhalf$ are dominated by observational uncertainties rather than theoretical uncertainties associated with $\beta(r)$. In both plots and in the inset the solid line indicates the one-to-one relation.
The stellar velocities used to derive the globular cluster (GC) masses (in conjunction with photometry from \citet{Harris_96}) were obtained from (lowest to highest mass): NGC 5053 \citep{YanCohen_96}, NGC 6171 \citep{Piatek_94}, NGC 288 \citep{Pryor_91}, NGC 104 \citep{Mayor_83}, NGC 362 \citep{Fischer_93}, NGC 5272 \citep{Pryor_88}, and NGC 2419 \citep{Baumgardt_09}. The kinematic data for the classical dSphs were taken from \citet{Munoz_05, Koch_07, Mateo_08, Walker_09a}, and data for the post-SDSS dSphs were taken from \citet[][]{Munoz_06, SimonGeha_07, Geha_09}, and Willman et al. (in preparation). The kinematic data for the ellipticals are as follows (from lowest to highest mass): NGC 185 \citep{DeRijcke_06}, NGC 855 \citep{SandP_00}, NGC 4478 \citep{SandP_97a}, NGC 731 \citep{SandP_00}, NGC 3853 \citep{SandP_97b}, and NGC 499 \citep{SandP_97c}. The photometric data for the MW dSphs, dEs, and ellipticals are referenced in Table 1. These specific dwarf ellipticals and ellipticals were chosen because they had extended kinematic data (to $\Rhalf$) and showed little rotation.
}
\label{fig:Mrhalf}
\end{figure*} 
%

\subsubsection{General velocity dispersion anisotropy}

Here we provide a qualitative understanding of why our 
mass estimator works well in the general $\beta(r)$ case.
We begin by reconsidering the derivation of 
$\ave{\sigmatot^2}$, now allowing $\beta$ to vary with radius.
It is clear that the peak in the integrand in Equation \ref{eq:sigtot}
will shift to a position where
$\gamma_\sigma+\gamma_\star+2\beta^\prime/(3-2\beta)=3$. Thus even if
$\gamma_\sigma$ is moderately small, the peak may be shifted due to
the third term. For small values of $\beta$, the typical 
$|\beta^\prime|/(3-2\beta)$ values are also small in our parameterizations 
(Equations \ref{eq:betaprofile} and \ref{eq:betaprofile2}) 
and hence the peak is close to $\rthree$ as in the
constant $\beta$ case. For large negative values of $\beta$, the peak
of the $\ave{\sigmatot^2}$  integrand is essentially at $\req$, but
this does not imply that $\req$ is close to $\rthree$. However, if
$\beta(\rthree)$ is not small, then $\beta^\prime(\rthree)$ is constrained by
Equation \ref{eq:sigmaprime}. This can be realized because the term that determines
the shift in the peak of Equation \ref{eq:forma} for large negative $\beta(\rthree)$
values is
\beq
\gamma_\sigma(\rthree)+\beta^\prime(\rthree)/(1-\beta(\rthree))  \propto 3\avelos(\rthree)-\sigmatot^2(\rthree).
\eeq
The simplest solution to this equation and Equation \ref{eq:sigmaprime}
which is consistent with the Jeans equation is $3\avelos
\simeq \sigmatot^2(\rthree)$ and $\req \simeq \rthree$. Our full mass likelihoods
derived from analyzing observed data confirm this expectation. 

Since we have argued that the mass enclosed within $\rthree$ should be
approximately independent of $\beta(r)$, we may now derive this mass
by simply using Equation \ref{eq:massjeans} with $\beta = 0$ at $r = \rthree$:  
\begin{eqnarray}
\label{eq:formula}
M(\rthree)  & = & \frac{\rthree\: \sigma_r^2(\rthree)}{G} \left. 
\left[ \gamma_\star(\rthree) + \gamma_\sigma(\rthree) \right ] \right|_{\beta =0} \nonumber\\
& \simeq & \left. \frac{3 \: \rthree \: \sigma_r^2(\rthree)}{G} \right|_{\beta =0} 
\simeq \frac{3 \: \rthree \: \ave{\sigmalos^2}}{G} .
\end{eqnarray}
This is again Equation \ref{eq:main} with $\rhalf \simeq \rthree$.
In the second line we are using the fact that $3\, \sigma_r^2 =
\sigmatot^2$ for $\beta=0$ and our result from the previous section
that $\sigmatot^2(\rthree) \simeq \ave{\sigmatot^2}$. 

It is worth emphasizing that the ideal radius for mass determination is $\rthree$ and not $\rhalf$.
As one moves away from $\rthree$, the uncertainty in $\beta(r)$ will start dominating over 
kinematic (or photometric) errors. However, typically the observational errors
on both $\rthree$ and $\avelos$ are large enough that the slight ($\sim 15\%$) difference between
$\rhalf$ and $\rthree$ will not matter. For this reason we have opted to present our results using the more familiar
deprojected half-light radius in what follows. We find that for constant $\beta$ or for our monotonically varying $\beta(r)$
form, both $M(\rhalf)$ and $M(\rthree)$ are equally well constrained by the data sets we consider 
when analyzing the population as a whole.

Of course, one expects the expression in Equation \ref{eq:main} to
fail in special cases. For example, if the line-of-sight velocity
dispersion declines very rapidly within the half-light radius (such
that $\gamma_\sigma \sim \gamma_\star$) then we would expect the
mass-anisotropy uncertainty to be minimized at a radius smaller than
$\rhalf$. However, if we ignore the very central regions of spheroids
with supermassive black holes, most dispersion-supported galaxies do
not show significant declines in their stellar velocity dispersion profiles 
within their half-light radii. Indeed, as we now discuss, we find that Equation
\ref{eq:formula} does a remarkably good job at reproducing the masses
for real galaxies that span a wide dynamic range in luminosity, size,
and mass -- at least under the assumption of spherical symmetry. 

\subsection{Tests}
\label{subsec:tests}

The left-hand panel of Figure \ref{fig:Mrhalf} presents the integrated
masses within $\rhalf$ as obtained using our fiducial likelihood analysis
for a variety of spheroidal systems plotted against the simple mass
estimator in Equation \ref{eq:main}. We see that this formula is
accurate over almost eight decades in $\Mhalf$. As detailed in the
caption, we use individual stellar velocity data sets in our
likelihoods for MW globular clusters and dSphs, and published velocity
dispersion profiles for the dwarf elliptical galaxies (dEs) and
elliptical galaxies (Es). Observed properties and derived masses for
each of these systems is presented in Table 1.    

To demonstrate the accuracy of the normalization in our formula we add
an inset into Figure \ref{fig:Mrhalf}, which zooms in to the region
populated by the so-called ``classical" (pre-SDSS) MW dSphs, since
they have the most well-measured and spatially extended stellar
velocity distributions and well-studied photometry. The dashed lines
indicate $\pm 10\%$ variation about the predicted relation. In the
right-hand panel of Figure \ref{fig:Mrhalf} we demonstrate that
Equation \ref{eq:main} also provides a good measure of 
uncertainties on $\Mhalf$ for the MW dSphs\footnote{Leo IV is not 
included in the right-hand panel because it is has very few 
accurate kinematic stellar measurements.} (compare to Figure
\ref{fig:Mlim}). The errors on the vertical axis are 68\% likelihoods
derived from our analysis, while the errors along the horizontal axis
are calculated by simply propagating the observational errors on
$\rhalf$ and $\sigmalos$ through Equation \ref{eq:main}. This rough
agreement is consistent with the $\Mhalf$ uncertainty being dominated
by observational errors as opposed to the uncertainty in $\beta$, as
expected. 

%
\begin{figure*}
\includegraphics[width=85mm]{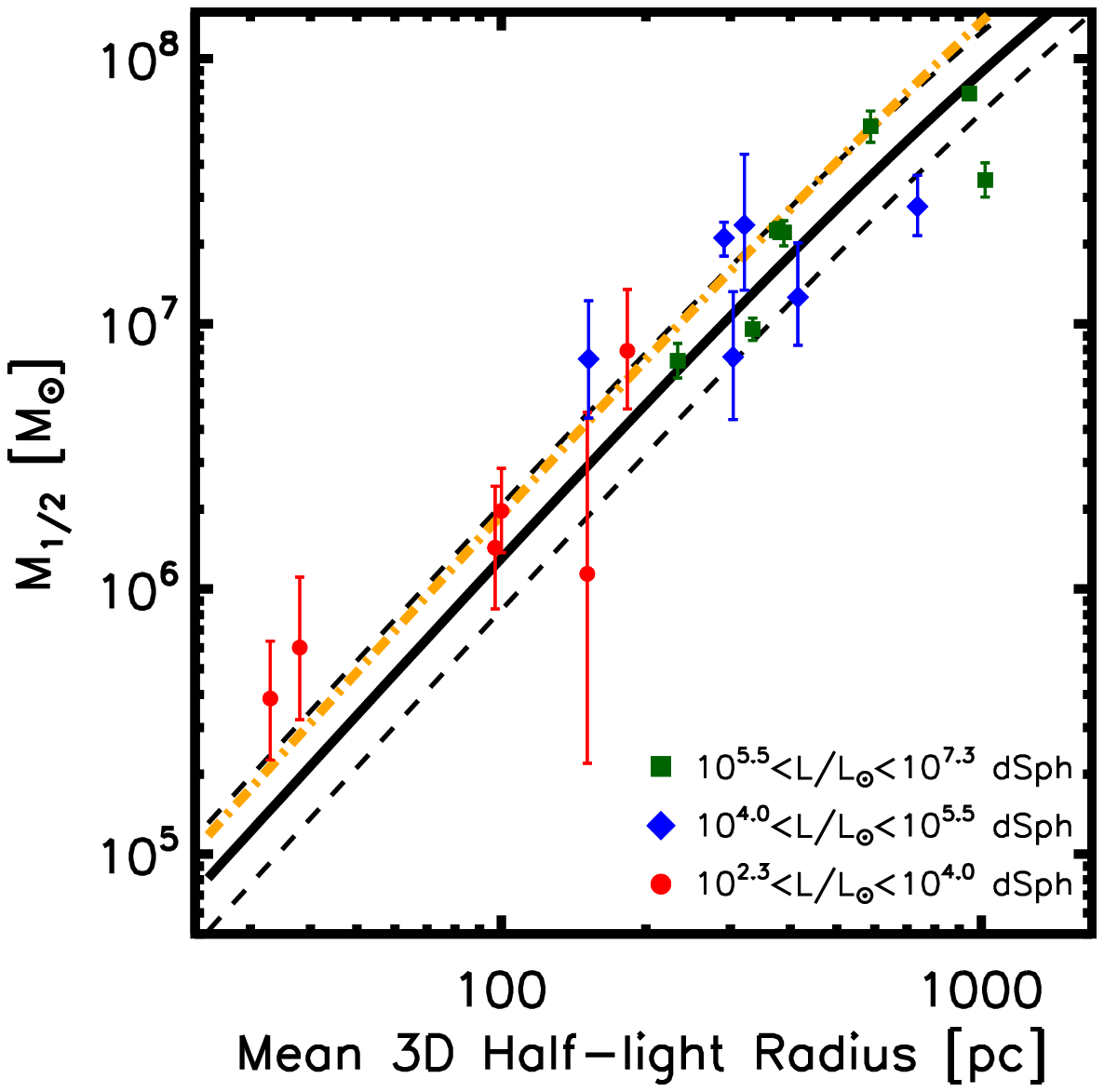}
\hspace{5mm}
\includegraphics[width=85mm]{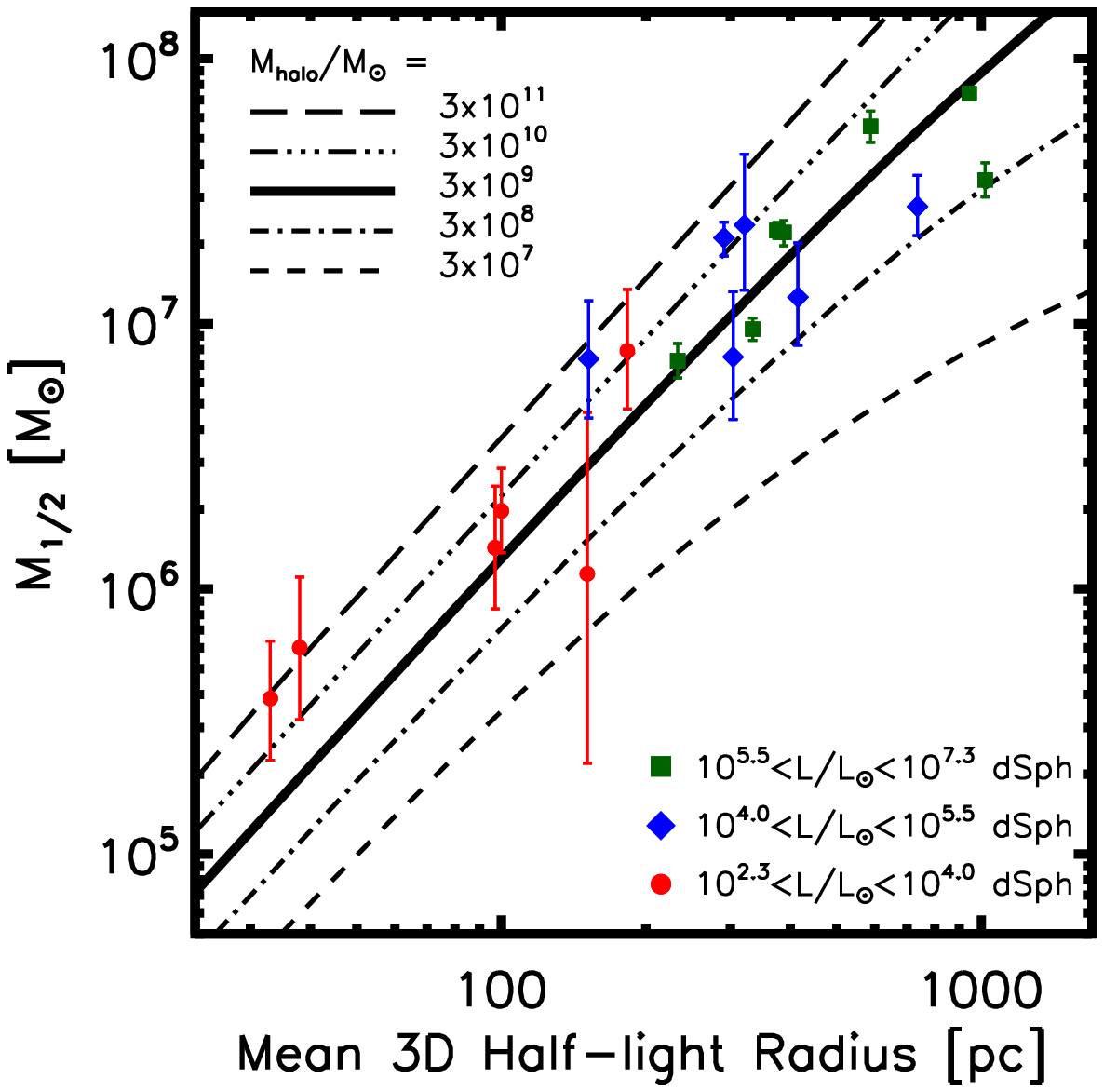}
\caption{The half-light masses of the Milky Way dSphs plotted against $\rhalf$. {\em Left:} The solid black line shows the NFW mass profile for a field halo of $\Mtwohun = 3\times10^9 \Msun$ at $z=0$ expected for a WMAP5 cosmology \citep[$c=11$ according to][]{Maccio_08}, where the two dashed lines correspond to a spread in concentration of $\Delta \log_{10}(c)=0.14$, as determined by N-body simulations \citep{Wechsler_02}. The orange dot-dashed line shows the profile for a median $\Mtwohun = 3 \times10^9 \Msun$ at $z=3$. 
{\em Right:} The same data points along with the (median c) NFW mass profiles for halos with $\Mtwohun$ masses ranging from $3 \times 10^7 \Msun$ to $3\times10^{11} \Msun$ (from bottom to top). We note that while all but one of the MW dSphs are consistent with sitting within a halo of a common mass (left), many of the dwarfs can also sit in halos of various masses (right). There is no indication that lower luminosity galaxies (red circles) are associated with less massive halos than the highest mass galaxies (green squares), as might be expected in simple models of galaxy formation. None of these galaxies are associated with a halo less massive than $\Mtwohun \simeq 3\times 10^8 \Msun$.
}
\label{fig:Mrhalfvsrhalf}
\end{figure*} 
%

It is worth emphasizing that
Equation \ref{eq:main} is not able to capture the full uncertainty
on the half-light mass in cases where the kinematic data does not
constrain $\sigmalos$ beyond $\Rhalf$. While our full likelihood
procedure naturally takes into account any limitations in the data and
factors them into the resultant mass uncertainty, Equation
\ref{eq:main} was derived under the assumption that $\sigmalos$
remains constant out beyond $R \simeq \Rhalf$. The lack of extended
kinematic data is manifest in the more massive galaxies
presented in Figure \ref{fig:Mrhalf}. A careful examination of the dEs
and regular Es (those with $\Mhalf > 10^8 \Msun$) reveals that the
errors on the ordinate axis are on average 0.05 dex larger than
the errors on the abscissa. Therefore, in cases where extended
kinematics are not available, if one is willing to assume that an
unmeasured velocity dispersion profile does not fall too sharply
within $\sim 1.5 \, \Rhalf$ (as is seen in most galaxies with measured
dispersion profiles that extend this far), then our proposed estimator
should provide an accurate description of the half-light mass and the associated
uncertainty (via simple error propagation). If one does not wish to accept the 
assumption of a flat $\sigmalos$ profile, then adding an error of 0.05 dex
 to the propagated mass error provides a reasonable means to allow for
 a range of $\beta$ profiles.

We note that all of the mass modeling presented so far
has been done by allowing $\beta(r)$ to vary according to
the profile in Equation \ref{eq:betaprofile}. This allows for
$\beta(r)$ to vary monotonically with three free parameters. 
All of the results quoted in Table 1 allow for this sort of 
spatial variation in $\beta(r)$. Though this profile is fairly general 
and has the added virtue that it is reminiscent of the anisotropy
of cold dark matter particles found in numerical
simulations \citep[e.g.,][]{carlberg:97}, we have also performed our 
analysis using the $\beta(r)$ form in Equation \ref{eq:betaprofile2}, which allows
for an extremum within the stellar light distribution.
We find that even with this unusual family of $\beta(r)$ profiles, 
no bias in the mass estimates exists (within either $\rthree$ or $\rhalf$) 
between the two $\beta(r)$ forms. However, the errors on $\Mhalf$
increased by roughly 0.05 dex when the (rather extreme) second
$\beta(r)$ form was used. The errors on $M(\rthree)$ were slightly less affected. 
Hence Equation \ref{eq:main2} becomes preferable to Equation \ref{eq:main} for the most
general $\beta(r)$ profiles, as long as the required photometric measurements (for $\rthree$) and
kinematic data sets (for $\avelos$) are good enough to warrant the need for $10 \%$ accuracy.

Before moving on, we mention that in Appendix \ref{sec:otherformulae} we perform a similar test using our full mass modeling procedure
against a popular mass estimator for dSphs known as the \citet{Illingworth_76} approximation.  We show that the Illingworth
formula fails both because it systematically under-predicts masses and
because it under-predicts mass uncertainties. The main reason for the failure
is that it was derived for mass-follows-light globular clusters using $\beta=0$. 
It was never intended to be generally applicable to dark-matter dominated systems like dSphs.

Lastly, in Appendices \ref{subsec:Spitzer} and \ref{subsec:C06} we compare Equation \ref{eq:main}
to the mass estimators presented by \citet{Spitzer_69} and \citet{Cappellari_06}.

\section{Discussion}
\label{sec:discussion}
%
We have shown that the integrated mass within the half-light radius of
spherically symmetric, dispersion-supported systems is very well
constrained by line-of-sight kinematic observations with only mild
assumptions about the spatial variation of the stellar velocity
dispersion anisotropy: $\Mhalf = 3 \, G^{-1} \, \avelos \, \rhalf$.
Mass determinations at larger {\em and smaller} radii are 
much more uncertain because of the uncertainty in $\beta(r)$. In the
following two subsections we use $\Mhalf$ determinations to examine
the dark matter halos of MW dSphs and to explore the mass-luminosity
relation in dispersion-supported galaxies as a function of mass
scale.

\subsection{Dwarf spheroidal satellite galaxies of the Milky Way}
As an example of the utility of $\Mhalf$ determinations, both panels
of Figure \ref{fig:Mrhalfvsrhalf} present $\Mhalf$ vs. $\rhalf$ for MW
dSph galaxies. We have used our full mass likelihood approach in
deriving these masses and associated error bars, though had we simply
used Equation \ref{eq:main} the result would have been very
similar. In interpreting this figure, it is important to emphasize
that the galaxies represented here span almost five orders of
magnitude in luminosity. Relevant parameters for each of the galaxies
are provided in Table 1. The symbol types labeled on the plot
correspond to three wide luminosity bins (following the same scheme
represented in Figure \ref{fig:Mrhalf}). Note that among galaxies with
the same half-light radii, there is no clear trend between luminosity
and density. We return to this noteworthy point below. 

It is interesting now to compare the data points in Figure
\ref{fig:Mrhalfvsrhalf} to the integrated mass profile $M(r)$
predicted for $\LCDM$ halos of a given $\Mtwohun$
mass. We define $\Mtwohun$ as the halo mass corresponding to an
overdensity of 200 compared to the critical density. In
the limit that dark matter halo mass profiles $M(r)$ map 
in a one-to-one way with their $\Mtwohun$ mass \citep{nfw}, then the
points on this figure may be used to estimate an associated halo mass
for each galaxy. The association is not perfect for three reasons: 1)
some scatter exists in halo concentration at fixed mass and redshift
\citep[e.g.,][]{Jing_00, Bullock_01}; 2) the mapping between $M(r)$
and $\Mtwohun$ evolves slightly with redshift \citep[e.g.,][]{Bullock_01};   
 and 3) the MW satellites all reside within subhalos, which tend to lose mass 
after accretion from the field \citep[see][]{Kazantzidis_04}.  
Nevertheless, we may still examine the median $M(r)$ dark matter halo
profile for a given $\Mtwohun$ in order to 
provide a reasonable estimate their progenitor halo masses prior to
accretion onto the Milky Way. 

The solid line in the left panel of Figure \ref{fig:Mrhalfvsrhalf}
shows the mass profile for a NFW \citep{nfw} dark matter halo at
$z=0$ with a halo mass  $\Mtwohun = 3 \times 10^9 \Msun$. We have used
the median concentration ($c=11$) predicted by the
\citet{Bullock_01} mass-concentration model  updated by
\citet{Maccio_08} for WMAP5 $\LCDM$ parameters. The dashed lines
indicate the expected $68 \%$ scatter about the median concentration
at this mass. The orange dot-dashed line shows the expected $M(r)$ profile
for the same mass halo at $z=3$ (corresponding to a concentration of $c=4$), 
which provides an estimate of the scatter that would result from the 
scatter in infall times. We see that each MW dSph is consistent with 
inhabiting a dark matter halo of mass $\sim 3 \times 10^9 \Msun$
\citep{Strigari_08}. \citet{Walker_09b} recently submitted an article
that presented a similar result for Milky Way dSphs by examining the
mass within a radius $r = \Rhalf$ rather than $r = \rhalf$ as we have
done. Note that since $\Rhalf \simeq 0.75\, \rhalf$, the mass within
$r=\Rhalf$ is still somewhat constrained without prior knowledge
of $\beta$. 

The right panel in Figure \ref{fig:Mrhalfvsrhalf} shows the same data
plotted along with the median mass profiles for several different halo
masses. Clearly, the data are also consistent with MW dSphs populating
dark matter halos of a wide range in $\Mtwohun$. As described in
\citet{Strigari_08}, there is a weak power-law relation between a
halo's inner mass and its total mass (e.g., $M(300 {\rm pc}) \propto
\Mtwohun^{1/3}$ at $\Mtwohun \simeq 10^9 \Msun$), and this makes a precise 
mapping between the two difficult. Nevertheless, several interesting 
trends are manifest in the comparison. 

First, all of the MW dSphs are associated with halos more massive than
$\Mtwohun \simeq 10^8 \Msun$. This provides a very stringent
limit on the fraction of the baryons converted to stars in these
systems. More importantly, there is no systematic relationship
between dSph luminosity and the $\Mtwohun$ mass profile that they
most closely intersect. The ultra-faint dSph population (red circles)
with $\LV <$ 10,000 $\Lsun$ is equally likely to be associated with
the more massive dark matter halos as are classical dSphs that are
more than 1,000 times brighter (green squares). Indeed, a naive
interpretation of the right-hand panel of Figure
\ref{fig:Mrhalfvsrhalf} shows that the two least luminous satellites
(which also have the smallest $\Mhalf$ and $\rhalf$ values) are
associated with halos that are either {\em more massive} than any of the
classical MW dSphs (green squares), or have abnormally large 
concentrations (reflecting earlier collapse times) for their halo mass.
This general behavior is difficult
to reproduce in models constructed to confront the Milky Way satellite
population \citep[e.g.,][]{Koposov_09, Li_09, Maccio_09, Munoz_09, 
Salvadori_09, Busha_10, Kravtsov_10}, which typically predict a noticeable trend
between halo infall mass and dSph luminosity. It is possible that we
are seeing evidence for a new scale in galaxy formation \citep{Strigari_08}
or that there is a systematic bias that makes less luminous
galaxies that sit within low-mass halos more difficult to detect than
their more massive counterparts \citep{Bovill_09, Bullock_09}. 

\begin{figure*}
\includegraphics[width=56mm]{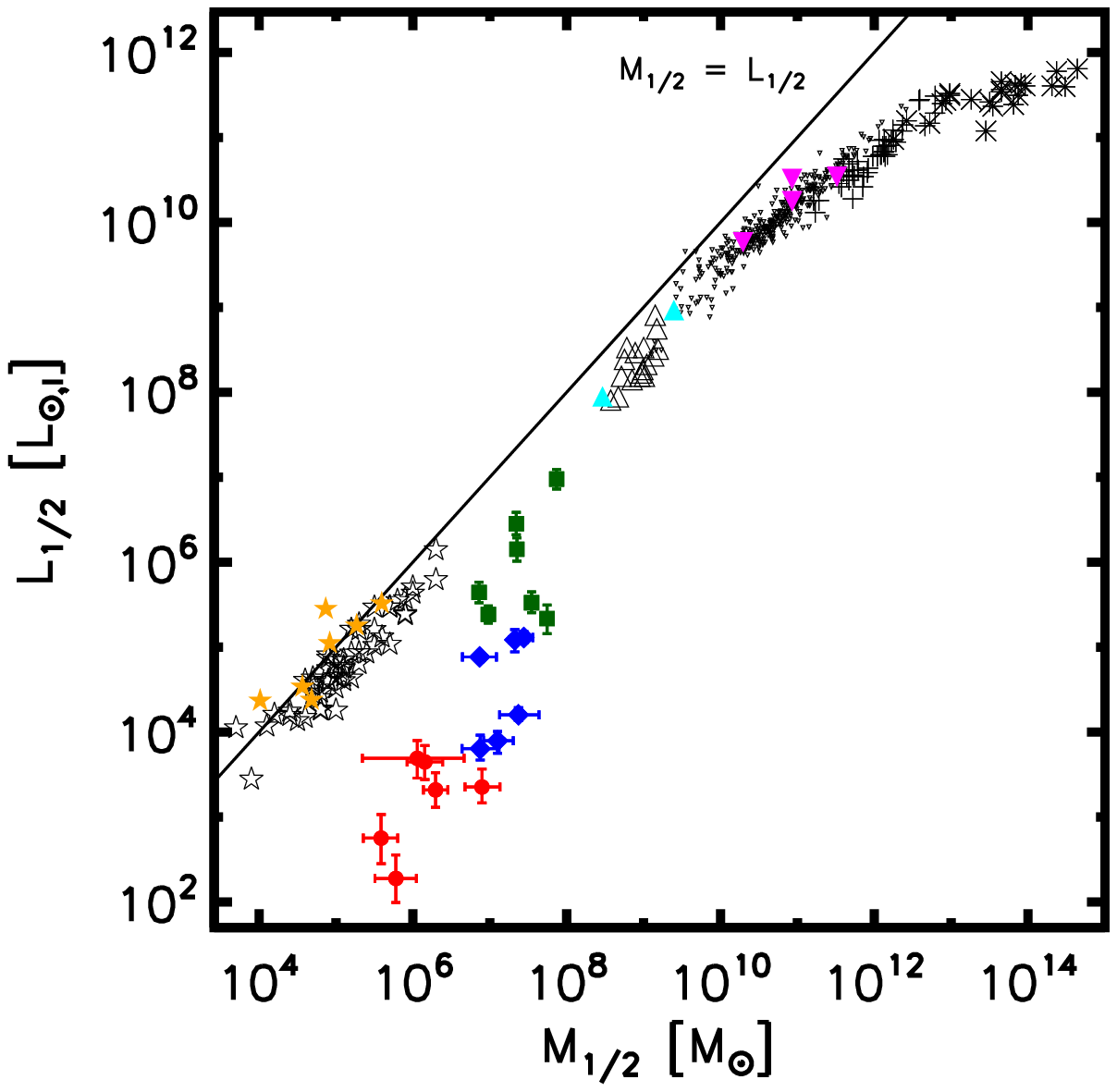}
\hspace{1mm}
\includegraphics[width=56mm]{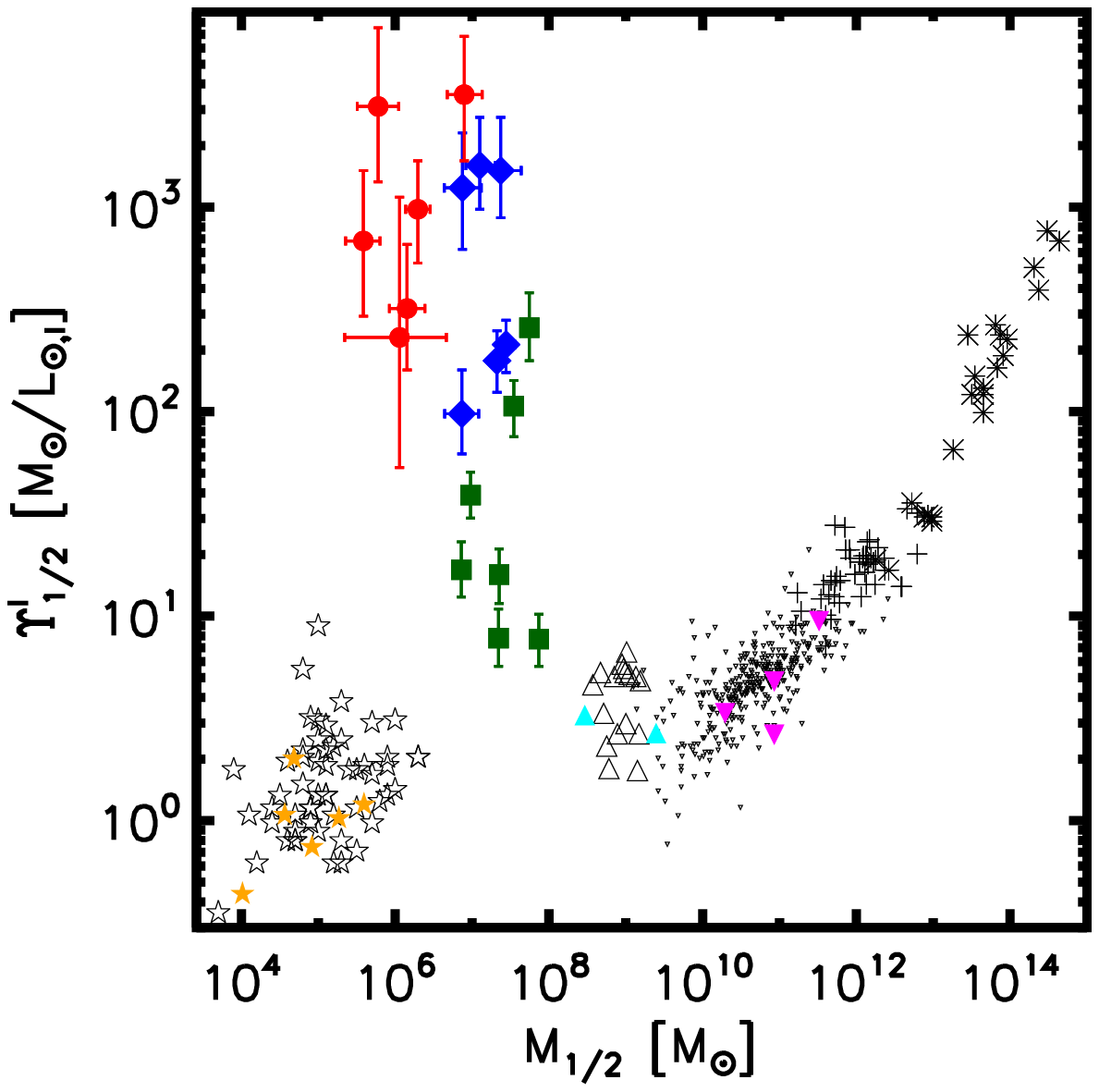}
\hspace{1mm}
\includegraphics[width=56mm]{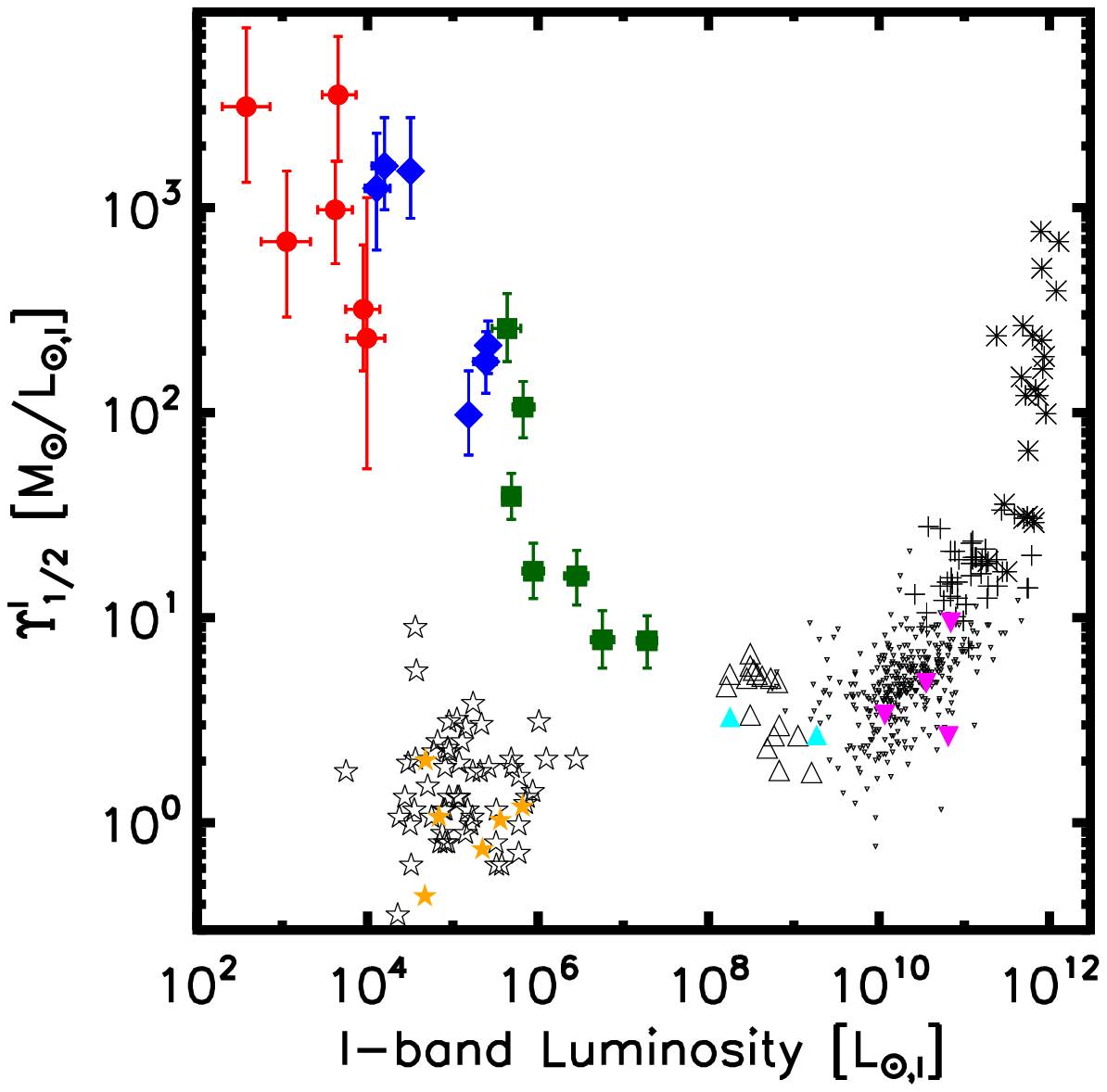}
\caption{{\em Left:} The half I-band luminosity $\Lhalf$ vs. half-light mass $\Mhalf$ for a broad population of spheroidal galaxies. {\em Middle:} The dynamical I-band half-light mass-to-light ratio $\moverl$ vs. $\Mhalf$ relation. {\em Right:} The equivalent $\moverl$ vs. total I-band luminosity L$_{\rm I} = 2 \, \Lhalf$ relation. The solid line in the left panel guides the eye with $\Mhalf = \Lhalf$ in solar units. The solid, colored points are all derived using our full mass likelihood analysis and their specific symbols/colors are linked to galaxy types as described in Figure \ref{fig:Mrhalf}. The I-band luminosities for the MW dSph and GC population were determined by adopting M92's $V - I = 0.88$. All open, black points are taken from the literature as follows. Those with $\Mhalf > 10^8 \Msun$ are modeled using Equation \ref{eq:main} with $\sigmalos$ and $\rhalf$ culled from the compilation of \citet{Zaritsky_06}: triangles for dwarf ellipticals \citep{Geha_03}, inverse triangles for ellipticals \citep{Jorgensen_96, MatkovicGuzman_05}, plus signs for brightest cluster galaxies \citep{OegerleHoessel_91}, and asterisks for cluster spheroids, which, following \citet{Zaritsky_06}, include the combination of the central brightest cluster galaxy and the extended intracluster light. Stars indicate globular clusters, with the subset of open, black stars taken from \citet{PryorMeylan_93}.}
\label{fig:manifold}
\end{figure*} 
%

\subsection{The global population of dispersion-supported stellar systems}
A second example of how accurate $\Mhalf$ determinations may be used
to constrain galaxy formation scenarios is presented in Figure
\ref{fig:manifold}, where we examine the relationship between the
half-light mass $\Mhalf$ and the half-light I-band luminosity $\Lhalf = 0.5
\, \LI$ for the full range of dispersion-supported stellar systems in the
Universe: globular clusters, dSphs, dwarf ellipticals, ellipticals,
brightest cluster galaxies, and extended cluster spheroids. Each
symbol type is matched to a galaxy type as detailed in the caption.
We provide three representations of the same information in order to
highlight different aspects of the relationships: $\Mhalf$
vs. $\Lhalf$ (left panel); the dynamical I-band mass-to-light ratio within 
the half-light radius $\moverl$ vs. $\Mhalf$ (middle panel); and
$\moverl$ vs. total I-band luminosity $\LI$ (right panel).  

Masses for the colored points are derived using our full mass
likelihood approach and follow the same color and symbol convention as
in Figure \ref{fig:Mrhalf}. All of the black points that represent
galaxies were modeled using Equation \ref{eq:main} with published
$\sigmalos$ and $\rhalf$ values from the literature.\footnote{The 
masses for the open, black stars (globular clusters) were taken directly
from \citet{PryorMeylan_93}.} The middle and right panels are
inspired by (and qualitatively consistent with) Figures 9 and 10 from
\citet{Zaritsky_06}, who presented estimated dynamical mass-to-light ratios 
as a function of $\sigmalos$ for spheroidal galaxies that spanned two 
orders of magnitude in $\sigmalos$.

We note that the asterisks in Figure \ref{fig:manifold} are cluster spheroids \citep{Zaritsky_06}, which are defined for any galaxy cluster to be the sum of the extended low-surface brightness intracluster light component and the brightest cluster galaxy's light. These two components  are difficult to disentangle, but the total light tends to be dominated by the intracluster piece. One might argue that the total cluster spheroid is more relevant than the brightest cluster galaxy because it allows one to compare the dominant stellar spheroids associated with individual dark matter halos over a very wide mass range self consistently. Had we included analogous diffuse light components around less massive galaxies (e.g., stellar halos around field ellipticals) the figure would change very little, because halo light is of minimal importance for the total luminosity in less massive systems \citep[see][]{Purcell_07}. One concern is that the central cluster spheroid mass estimates here suffer from a potential systematic bias because they rely on the measured velocity dispersion of cluster galaxies for $\sigmalos$ rather than the velocity dispersion of the cluster spheroid itself, which is very hard to measure \citep{Zaritsky_06}.\footnote{In addition, concerns exist with the assumption of dynamical equilibrium. However, \citet{Willman_04} demonstrated with a simulation that using the intracluster stars as tracers of cluster mass is accurate to $\sim$ 10\%.} For completeness, we have included brightest cluster galaxies on this diagram (plus signs) and they tend to smoothly fill in the region between large elliptical galaxies (inverse triangles) and the cluster spheroids (asterisks).

There are several noteworthy aspects to Figure \ref{fig:manifold}, which are each highlighted in a slightly different fashion in the three panels. First, as seen most clearly in the middle and right panels, the dynamical half-light mass-to-light ratios of spheroidal galaxies in the Universe demonstrate a minimum at $\moverl \simeq 2-4$ that spans a remarkably broad range of masses $\Mhalf \simeq 
10^{9-11} \Msun$ and luminosities $\LI \simeq 10^{8.5-10.5} \, \Lsun$.   It is interesting to note the offset in the average dynamical mass-to-light ratios between globular clusters and L$_\star$ ellipticals, which may suggest that even within $\rhalf$, dark matter may constitute the majority of the mass content of L$_\star$ elliptical galaxies.  Nevertheless, it seems that dark matter plays a clearly dominant dynamical role ($\moverl \gtrsim 5$) within $\rhalf$ in only the most extreme systems \citep[see similar results by][who study slightly more limited ranges of spheroidal galaxy luminosities]{Dabringhausen_08, Forbes_08}. The dramatic increase in dynamical half-light mass-to-light ratios at both smaller and larger mass and luminosity scales is indicative of a decrease in the efficiency of galaxy formation in the smallest and largest dark matter halos.  It is worth mentioning that a qualitatively similar trend in the relationship between $\Mtwohun$ and L must exist if $\LCDM$ is to explain
 the luminosity function of galaxies \citep[e.g.,][]{WhiteRees_78, MarinoniHudson_02, Yang_03,CR_09, Moster_10}.  While the relationship presented in Figure \ref{fig:manifold} focuses on a different mass variable, the similarity in the two relationships is striking, and generally encouraging for the theory.

One may gain some qualitative insight into the physical processes that drive galaxy formation inefficiency in faint vs. bright systems by considering the $\Mhalf$ vs. $\Lhalf$ relation (left panel) in more detail.  We observe three distinct power-law regimes $\Mhalf \propto \Lhalf^\xi$ with $\xi <1$, $\xi \simeq 1$, and $\xi > 1$ as mass increases. Over the broad middle range of galaxy masses, $\Mhalf \simeq 10^{7-11} \Msun$, mass and light track each other quite closely with $\xi \simeq 1$, while very faint galaxies obey $\xi \simeq 1/2$, and bright elliptical galaxies have $\xi \simeq 4/3$ transitioning to $\xi \gg 1$ for the most luminous cluster spheroids. One may interpret the transition from $\xi < 1$ in faint galaxies to $\xi > 1$ in bright galaxies as a transition between mass-suppressed galaxy formation to luminosity-suppressed galaxy formation. That is, for faint galaxies ($\xi < 1$), we do not see any evidence for a low-luminosity threshold in galaxy formation, but rather we are seeing behavior closer to a threshold (minimum) mass with variable luminosity. For brighter spheroids with $\xi > 1$, the increased dynamical mass-to-light ratios are driven more by increasing the mass at fixed luminosity, suggestive of a maximum luminosity scale. 

Regardless of the interpretation of Figure \ref{fig:manifold}, it provides a useful empirical benchmark against which theoretical models can be compared. Interestingly, two of the least luminous dSph satellites of the Milky Way have the highest dynamical mass-to-light ratios $\moverl \simeq$ 3,200 of any collapsed structures shown, including intra-cluster light spheroids, which reach values of $\moverl \simeq 800$. It is well known that the ultra-faint dSphs are the most dark matter dominated objects known \citep[e.g.,][]{Strigari_08}. For example, they have much lower baryon-to-dark matter fractions $f_{b} \sim \Omega_{b}/\Omega_{dm} \lesssim 10^{-3}$ than galaxy clusters $f_{b} \simeq 0.1$. Now we see that ultra-faint dSphs also have higher dynamical mass-to-visible light ratios within their stellar extents than even the (well-studied) galaxy cluster spheroids.

\section{Conclusions}

We have shown that line-of-sight kinematic observations enable
accurate mass determinations for spherical, dispersion-supported
galaxies within a characteristic radius that is approximately equal to
$\rthree$, the radius where the log-slope of the stellar density profile
is $-3$. For a wide range of observed spheroidal galaxy stellar
luminosity profiles $\rthree$ is close to the 3D deprojected half-light radius
$\rhalf$, and we have opted to quote our main result in terms of the
mass enclosed within $\rhalf$.  While mass determinations at both
larger {\em and smaller} radii remain uncertain because of the
unknown velocity anisotropy (\S 3.1), the half-light mass is
accurately determined by the simple expression $\Mhalf = 3 \, G^{-1} \, 
\avelos \, \rhalf \simeq 4 \, G^{-1} \, \avelos \, \Rhalf$ as long as the
velocity dispersion profile $\sigmalos(R)$ remains relatively flat out
to the 2D projected half-light radius $\Rhalf$. We derived this
expression analytically using a few observationally-motivated
assumptions in \S 3.2, and demonstrated its accuracy over eight orders
of magnitude in both luminosity and in $\Mhalf$ by comparing it to detailed modeling of
real galaxy data in \S 3.3. The two main assumptions we have made in this
work are that the systems that we are analyzing are spherically
symmetric and are in dynamical equilibrium. Testing the accuracy of 
Equation \ref{eq:main} as a function of ellipticity will be an important future step.  

As an example of the usefulness of the $\Mhalf$ estimator, we applied
our result to the dSph satellite population of the Milky Way and
specifically used the observed $\Mhalf$ vs $\rhalf$ relation to
associate a dark matter halo $\Mtwohun$ mass to each galaxy. By
allowing for the expected scatter in halo concentrations at fixed
mass, we showed that all of the MW dSphs are consistent with
inhabiting dark matter halos of mass $\Mtwohun \simeq 3 \times 10^9
\Msun$. We also showed that a range of $\Mtwohun$ values from $\sim 
10^8 \Msun$ to $3 \times 10^{11} \Msun$ is allowable as well, but
that no trend exists between the associated $\Mtwohun$ and galaxy
luminosity, despite the fact that these galaxies span over
four orders of magnitude in luminosity. Specifically, the lowest
luminosity dSphs ($\LV \simeq 500 \Lsun$) are at least as dense as, if not
more dense than, the brightest MW dSphs ($\LV \simeq 10^7 \Lsun$) when
normalized against the inner power-law mass profiles expected in
$\LCDM$ halos. This last point is difficult to reproduce in
models that assume a monotonic mapping between $\Mtwohun$ and galaxy
luminosity. It is worth emphasizing that none of the MW dSphs are
associated with dark matter halos smaller than $\Mtwohun \simeq 
10^8 \Msun$, and this alone provides a very tight constraint on
the fraction of baryons converted to stars in these systems. Of
course, these results assume that no systematic biases in
the kinematic data for dSph galaxies are present. One particular 
worry is the effect of binary stars. \citet{Minor_10} estimate that
medium-to-high binary fractions can inflate velocity dispersions by up to
$\sim 20 \%$ in the smallest dSphs. This will have to be taken into account in
future work, at least for the classical dwarfs that only have $\sim 10\%$ errors
on their $\Mhalf$ estimates.

We went on to explore the relationship between $\Mhalf$ and $\LI$ in
dispersion-supported galaxies, spanning the full range in I-band luminosity
and mass from globular clusters ($\LI \simeq 10^5 \Lsun$) to intra-cluster
light spheroids ($\LI \simeq 10^{12} \Lsun$). Globular clusters excluded,
the $\moverlI$ vs. $\Mhalf$ relation for dispersion-supported galaxies
follows a U-shape, with a broad minimum near $\moverlI \simeq 3$ that
spans dwarf elliptical galaxies to normal elliptical galaxies, a steep
rise to $\moverlI \simeq$ 3,200 for ultra-faint dSphs, and a more
shallow rise to $\moverlI \simeq 20$ for brightest cluster
ellipticals. If we include intra-cluster light spheroids in the
analysis, the rise continues to $\moverlI \simeq 800$ for the largest
galaxy clusters.  

Lastly, we note that Equation \ref{eq:main} can be 
rewritten succinctly in terms of the circular velocity at $\rhalf$ as 
\begin{equation}
{\rm V}_{\rm circ} \, (\, \rhalf \,) \, = \, \sqrt{3 \, \avelos}. 
\end{equation}
It is clear then that the maximum circular velocity of the dark matter
halo hosting such a dispersion-supported galaxy must obey 
${\rm V}_{\rm max} \, \ge  \, \sqrt{3 \, \avelos}$.

In summary, we have shown that the dynamical mass within the
deprojected half-light radius of dispersion-supported galaxies can be
measured accurately with only line-of-sight stellar velocity measurements.
We have provided a simple formula that allows this mass to be computed
given the measured luminosity-weighted square of the line-of-sight velocity
dispersion and the half-light radius. This result opens up new opportunities
to explore the relationships between stellar properties and the masses
of galaxies spanning approximately ten orders of magnitude in luminosity.\\

Acknowledgements $-$ We would like to acknowledge the referee, Gary Mamon, 
for a very careful reading of the manuscript which led to many improvements. 
In addition, we would like to thank Michele Cappellari, St{\'e}phane Courteau, 
Aaron Dutton, Louie Strigari, and Beth Willman for many in-depth discussions 
that led to improvements of this paper. We would also like to thank Jenny 
Graves for assistance in modeling the photometric properties of some of the 
elliptical galaxies. Lastly, we would like to thank David Buote, Oleg Gnedin, 
Evan Kirby, Hans-Walter Rix, Erik Tollerud, Scott Tremaine, and Simon White
for additional useful discussions.

\appendix
\onecolumn
\section{An expression for mass as a function of observables}
\label{ap:massLOS}

Here we derive a single expression for the mass profile of spheroidal galaxies $M(r;\beta)$ as a function of the observable combination $\Sigma_\star \sigmalos^2(R)$.

We begin by manipulating the standard equation for $\sigmalos$ in order to isolate the $R$ dependence into an integral kernel:
\begin{eqnarray}
\label{eq:LOSrelation2}
\Sigma_\star \sigmalos^2 (R) & = & \int^{\infty}_{R^2} n_\star \sigma_r^2(r) \left[1 - \frac{R^2}{r^2}\beta(r)\right] \frac{\rmd r^2}{\sqrt{r^2 - R^2}} \hspace{.4in}
\\ \nonumber
& = & \int_{R^2}^{\infty} \frac{n_\star \sigma_r^2}{r^2}\frac{(1 - \beta)r^2 + \beta(r^2 - R^2)}{\sqrt{r^2-R^2}} \rmd r^2 \\ \nonumber
& = & \int_{R^2}^{\infty} \frac{n_\star \sigma_r^2(1 - \beta)}{\sqrt{r^2-R^2}} \rmd r^2
- \left.\left(\sqrt{r^2-R^2} \int_{r^2}^\infty \frac{\beta n_\star \sigma_r^2}{\tilde{r}^2} \rmd \tilde{r}^2\right)\right|^\infty_{R^2}
+ \int_{R^2}^\infty\left(\int_{r^2}^\infty \frac{\beta n_\star \sigma_r^2}{\tilde{r}^2} \rmd \tilde{r}^2\right)\frac{1}{2}\frac{\rmd r^2}{\sqrt{r^2-R^2}} \\ \nonumber
& = & \int_{R^2}^{\infty} \left[\frac{n_\star \sigma_r^2}{(1 - \beta)^{-1}} + \int_{r^2}^\infty \frac{\beta n_\star \sigma_r^2}{2\tilde{r}^2} \rmd \tilde{r}^2 \right] \frac{\rmd r^2}{\sqrt{r^2-R^2}} \, , 
\end{eqnarray}
where we employed an integration by parts to achieve the third equality. Note that we have set the 
middle term on the third line to zero by making the physically-motivated assumption that
the combination $\beta n_\star \sigma_r^2$ falls faster than $r^{-1}$ at large r.

With this crucial manipulation in place, we may now utilize the following Abel inversion
\beq
\label{eq:abel}
f(x) = \int_x^\infty \frac{g(t) \rmd t}{\sqrt{t-x}} \Rightarrow 
g(t) = -\frac{1}{\pi}\int_t^\infty \frac{\rmd f}{\rmd x} \frac{\rmd x}{\sqrt{x-t}}
\eeq
to solve for
\beq
\label{eq:g1}
g(r^2) = n_\star \sigma_r^2 ( 1 - \beta ) +  \int_{\ln r}^\infty \beta n_\star \sigma_r^2 \rmd \ln \tilde{r}
\eeq
in terms of the observable combination $f(R^2) = \Sigma_\star \, \sigmalos^2(R^2)$:
\beq
\label{eq:g2}
g(r^2) = -\frac{1}{\pi}\int_{r^2}^\infty \frac{\rmd (\Sigma_\star \, \sigmalos^2)}{\rmd R^2} \frac{\rmd R^2}{\sqrt{R^2-r^2}} \, .
\eeq

In order to isolate $n_\star \sigma_r^2$ we equate Equations \ref{eq:g1} and \ref{eq:g2}, and then 
differentiate the resulting expression with respect to $\ln r$ (denoted by $^\prime$)
\beq
\label{eq:sigmaprime2}
\frac{(n_\star \sigma_r^2)^\prime}{(1-\beta)^{-1}} - \left(n_\star
  \sigma_r^2\right) \left(\beta + \beta^\prime\right) 
= -\frac{2 r^2}{\pi} \int_{r^2}^\infty \frac{\rmd^2 (\Sigma_\star \sigmalos^2)}{(\rmd R^2)^2}
\frac{\rmd R^2}{\sqrt{R^2 - r^2}} 
\eeq
and employ the integrating factor
\beq
h(r)=\exp\left[-\int_{\ln a}^{\ln r} \frac{\beta + \beta^\prime}{1-\beta} \rmd \ln \tilde{r}\right]
\eeq
with the constant $a$ chosen such that the value of the integrand goes to zero at the lower limit:

\beq
\label{eq:rhostar}
n_\star \sigma_r^2 (r; \beta)
= \frac{h^{-1}}{\pi}\int_{r^2}^{\infty}
\left[\int_{\tilde{r}^2}^{\infty} \frac{\rmd^2 (\Sigma_\star \sigmalos^2)}{(\rmd R^2)^2} \frac{\rmd R^2}{\sqrt{R^2 - \tilde{r}^2}}\right]
\frac{h \: \rmd \tilde{r}^2}{\beta-1}
 = \frac{h^{-1}}{\pi}\int_{r^2}^{\infty}\left[\int_{r^2}^{R^2} \frac{h}{\beta-1}
\frac{\rmd \tilde{r}^2}{\sqrt{R^2 - \tilde{r}^2}}\right]\frac{\rmd^2 (\Sigma_\star \sigmalos^2)}{(\rmd R^2)^2} \rmd R^2.
\eeq

If one wishes to adopt a parametric form for $\beta(r)$, $n_\star \,
\sigma_r^2$ can be determined using Equation \ref{eq:rhostar}, and
then inserted into the Jeans equation to find the cumulative mass
profile.\footnote{In the final stages of this work, we learned of an 
alternative derivation performed by \citet{Mamon_10} who provide single 
integral expressions for both constant anisotropy and special cases of 
$\beta(r)$.} Note that
nothing guarantees a physical mass profile (i.e., the mass never decreases); 
given a very large number of stellar velocities with very low measurement
errors, one can restrict the anisotropy such that a physical mass is derived. 

If $\beta(r)$ is assumed to be constant, then the inner integral of the 
right-hand side of Equation \ref{eq:rhostar} can be written in terms 
of the incomplete Beta function: 
\beq
B_x(p, q) \equiv \int_0^x y^{p-1}(1-y)^{q-1} \rmd y.
\eeq
By utilizing the substitution $u=1-r^2/R^2$, we find
\beq
\label{eq:rhostaralt}
n_\star \sigma_r^2 (r; \beta) = \frac{r^{\:\beta/(1-\beta)}}{\pi(\beta-1)} 
\int_{r^2}^{\infty}R^{\tfrac{1-2\beta}{1-\beta}} B_{1-\tfrac{r^2}{R^2}}\left(\tfrac{1}{2}, \tfrac{2-3\beta}{2(1-\beta)}\right)
\frac{\rmd^2 (\Sigma_\star \sigmalos^2)}{(\rmd R^2)^2} \rmd R^2.
\eeq
By solving the Jeans equation we can derive the mass by first taking a derivative of Equation \ref{eq:rhostar}, and then inserting the form derived in Equation \ref{eq:rhostaralt}:
\beq
\label{eq:mass}
M(r; \beta) = \frac{1}{G\pi(\beta-1)n_\star(r)} 
\int_{r^2}^{\infty} R^2 \frac{\rmd^2 (\Sigma_\star \sigmalos^2)}{(\rmd R^2)^2} K(r, R; \beta) \, \rmd R^2
\eeq
where
\beq
K(r, R; \beta) = \frac{2 r^3/R^3}{\sqrt{1 - r^2/R^2}} 
+  \beta\frac{3 - 2\beta}{\beta-1}\left(\frac{r}{R}\right)^{\tfrac{1}{(1-\beta)}} 
B_{1 - \tfrac{r^2}{R^2}}\left(\tfrac{1}{2}, \tfrac{2-3\beta}{2(1-\beta)}\right).
\eeq

With this relation we have replaced the dependence of deriving the
mass of a dispersion-supported system from the unknown radial
dispersion $\sigma_r(r)$ with the second derivative of the observable
combination $\Sigma_\star \, \sigmalos^2(R)$. Note that determining the {\em slope} 
of the mass profile will require an additional derivative,
and thus we will require extremely accurate observational constraints
on both the light profile and the line-of-sight velocity
dispersion. We conjecture that the data will need to be so precise
that the assumption of spherical symmetry will no longer do the data
proper justice, and thus new derivations must be explored. 

\section{Useful conversions from 2D to 3D half-light radii}
\label{ap:conversions}

In this Appendix we present scaling relations to derive the 3D
deprojected half-light radius $\rhalf$ from the observed 2D projected
half-light radius $\Rhalf$ for several commonly used stellar
distributions. For the \citet{King_62} profile, $\rzero = \rcore$ and $c_k
\equiv \log_{10}(\rlimit/\rcore)$. A S\'ersic profile is defined as 
$I(R) = I(0) e^{-b_n \, (R/\rzero)^{1/n}}$, 
where $b_n$ is chosen such that $\rzero \equiv \Rhalf$. 
Note that although the exponential and Gaussian profiles are special 
cases of the n=1 and n=0.5 S\'ersic profiles, the $\Rhalf/\rzero$ relations 
are different due to the definitions of their scale radii: an exponential profile
is defined as $I(R) = I(0) e^{-R/\rzero}$ and a Gaussian profile is defined 
as $I(R) = I(0) e^{-R^2/2\rzero^2}$.\\ 
\\
\begin{tabular}{cccc}
\label{tab:Rhalf}
Profile& $\Rhalf/\rzero$ & $\rhalf/\Rhalf$ & $\rthree/\rhalf$ \\ \hline \hline
Exponential       & 1.678 & 1.329 & 1.15\\
Gaussian          & 1.178 & 1.307 & 1.13\\
King ($c_k$=0.70) & 1.185 & 1.322 & 1.13\\
Plummer           & 1.000 & 1.305 & 0.94\\
S\'ersic (n=2)    & 1.000 & 1.342 & 1.16\\
S\'ersic (n=4)    & 1.000 & 1.349 & 1.17\\
S\'ersic (n=8)    & 1.000 & 1.352 & 1.18\\
\end{tabular}\\
\\

We do not include the NFW profile due to the fact that the mass is divergent (thus, $\rhalf \rightarrow \infty$).
We also do note include the \citet{Einasto_65} profile in this table because it does not
well represent the baryonic tracer number density of most galaxies. The 
\citet{Hernquist_90} profile is sometimes used in place of a S\'ersic profile due to the 
ease of analytic manipulation. But we caution, as was pointed out in the original
paper, that the projected central surface 
brightness diverges logarithmically. This can cause $\gamma_\sigma$ to be quite 
large in magnitude, thus affecting the solution to Equation \ref{eq:req} more profoundly
than if the more well-behaved S\'ersic profile is used to model a tracer population.
 
Returning to the relations presented in the above table, for a King profile
\beq
\Rhalf/\rzero = 0.5439 + 0.1044c_k + 1.5618c_k^2 - 0.7559c_k^3 + 0.2572c_k^4
\eeq
to better than 2\% accuracy for $0.30 \leq c_k \leq 3.00$, and to better than 1\% accuracy for $0.40 \leq c_k \leq 3.00$. Also,
\beq
\rhalf/\Rhalf = 1.3088 + 0.0159c_k + 0.0066c_k^2 - 0.0035c_k^3 + 0.0004c_k^4
\eeq
to better than 0.04\% accuracy for $0.30 \leq c_k \leq 3.00$. Thus, the dominant error is in the relation between $\Rhalf$ and $\rzero$.

In regard to the family of S\'ersic profiles, as stated above, $\rzero \equiv \Rhalf$. To relate $\rhalf$ to $\Rhalf$, we utilize the following fit, which \citet{LimaNeto_99} state is valid to 0.25\% accuracy after testing against the numerical integration of the family of S\'ersic profiles corresponding to $0.10 \leq n^{-1} \leq 2.0$:
\beq
\rhalf/\Rhalf = 1.3560 - 0.0293 n^{-1} + 0.0023 n^{-2}.
\eeq

Thus, $\rhalf/\Rhalf \simeq 4/3$ is accurate to better than 2\% for most surface brightness profiles used to describe observed stellar systems. We also note that this result has been shown before for a wide range of S\'ersic profiles in \citet{Ciotti_91} and for the \citet{Plummer_11} profile in \citet{Spitzer_87}.

\section{Comparison With Other Mass Approximation Formulae}
\label{sec:otherformulae}
\subsection{Illingworth Formula}
%
\begin{figure*}
\includegraphics[width=85mm]{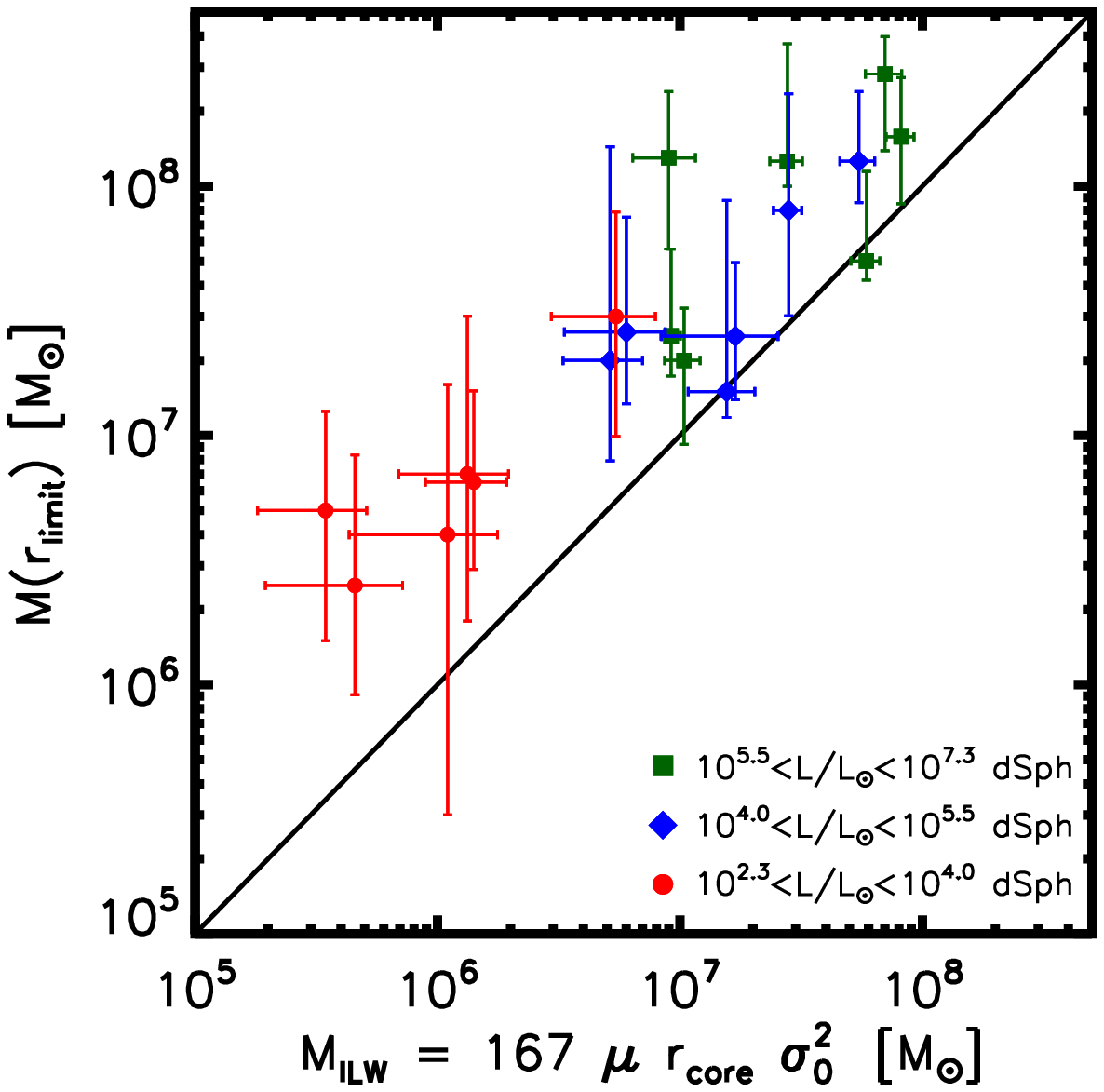}
\hspace{5mm}
\includegraphics[width=85mm]{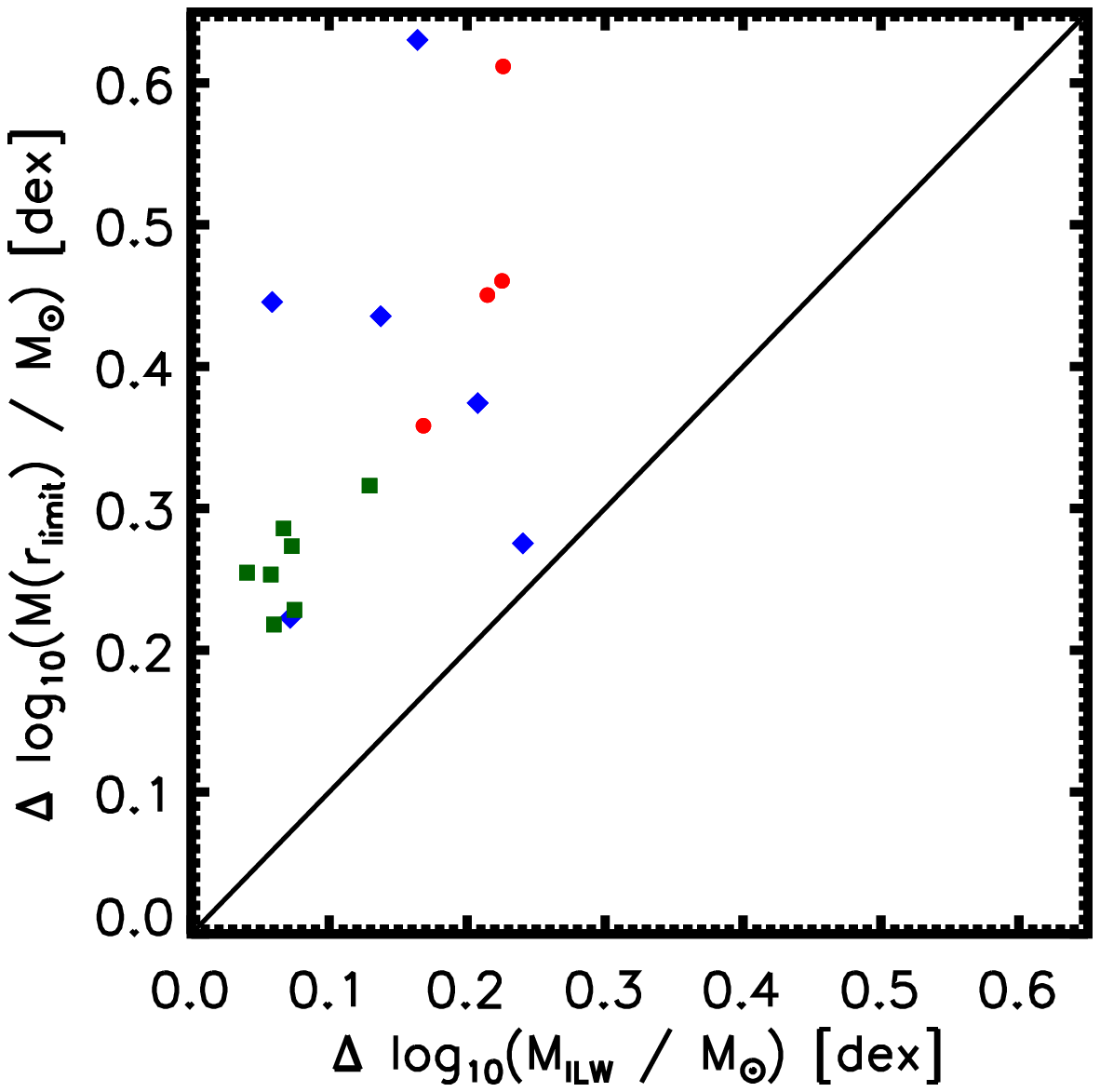}
\caption{
{\em Left:} The masses within the stellar extent for Milky Way dSphs. The vertical axis shows masses derived using individual stellar kinematics with our full likelihood procedure (see text) and the horizontal axis shows the ``Illingworth approximation", which is routinely used in the literature as a mass estimate for dSphs. 
{\em Right:} Errors on these masses for Milky Way dSphs. The vertical axis shows the 68\% error width derived from our full likelihood analysis and the horizontal axis shows the error width calculated by straightforward error propagation using Equation \ref{eq:Illingworth}. Note that this approximation tends to underestimate masses by up to an order of magnitude (left) and also under-estimates the relative error on the mass significantly (right). In both plots the solid line indicates the one-to-one relation.
}
\label{fig:Mlim}
\end{figure*}
%
Due to the large amount of attention that dSphs have received since new discoveries \citep{willman05a,willman05b,zucker06a,zucker06b, belokurov06,belokurov07,sakamoto06, irwin07, walsh07} were found in the public data releases of the SDSS \citep{York_00}, we will discuss an estimator that is often used to determine their masses. Because many dSphs look like larger versions of low-concentration globular clusters, the Illingworth formula (derived by \citet{Illingworth_76} for application only to globular clusters) is often used to estimate the masses of dSphs \citep[e.g.,][]{SeitzerFrogel_85, Suntzeff_93, Hargreaves_94, Mateo_98, SimonGeha_07}. Two explicit assumptions made by this formula are that the stellar velocity dispersion is isotropic and that the mass distribution follows a \citet{King_66} light distribution. Under these assumptions, the total mass within the stellar extent $\rlimit$ is stated as
\begin{equation}
\label{eq:Illingworth}
{\rm M}_{\rm Ilw} = 167 \: \mu \: \rcore \: \sigma_0^2 \: G^{-1},
\end{equation}
where $\sigma_0$ is the central line-of-sight velocity dispersion of the system, $\rcore$ is the King core radius and $\mu$ is a parameter that depends on the King concentration, $c_k \equiv \log_{10}(\rlimit / \rcore)$. It is common in the literature to set $\mu=8$ (incorrectly) for all dSphs based on a rough estimate provided in Table 4 of \citet{Mateo_98}. By adopting a value of $\mu$ without any error, many published mass uncertainties for dSphs do not properly include light profile uncertainties, which are typically only factored in from the error on $\rcore$. More important, however, is the implicit assumption that mass follows light in this formulation. While this is a reasonable assumption for globular cluster systems, the majority of the mass in dwarf galaxies does not necessarily follow the shape of their baryonic tracers \citep[e.g.,][]{SofueRubin_01, Walker_07, PNM_08}, as they are likely to be deeply embedded inside of dark matter halos \citep[e.g.,][]{WhiteRees_78}.
 
The left panel of Figure \ref{fig:Mlim} compares the masses M$(\rlimit)$ of Milky Way dSphs derived using our general approach to
Equation \ref{eq:Illingworth}. Symbol types correspond to luminosity bins, as indicated. For the general mass likelihoods, we analyze the kinematics of individual stars \citep[][Willman et al., in preparation]{Munoz_05, Munoz_06, Koch_07, Martin_07, SimonGeha_07, Mateo_08, Walker_09a, Geha_09}\footnote{We only accept stars whose projected distances lie within the lower limit of $\rlimit$ (see Table 1). For kinematics with assigned membership probabilities, we only accept those with p $\ge$ 0.9.}, in conjunction with the distances and stellar surface density profile parameters listed in Table 1. For the Illingworth approximation, we use the same observational datasets to calculate $\sigma_0$ (which is very close to the luminosity-weighted dispersion since the dispersion profiles for the MW dSphs are nearly constant with radius) and we follow the common practice of setting $\mu = 8$. Clearly, M$_{\rm Ilw}$ systematically underestimates the mass with this value of $\mu$.
This systematic difference follows from the fact that M$_{\rm Ilw}$ forces the mass profile to truncate at $\rlimit$ while the data prefer models where the mass distribution continues beyond the stellar extent.

However, the most dramatic difference between the full mass likelihoods and the Illingworth approximation is in the implied uncertainty. Errors on the vertical axis represent the 68\% width from the median of our derived mass likelihoods, while the symbol placement is indicative of the median of the likelihood. The errors on the horizontal axis propagate the observational errors on $\rcore$ and $\sigma_0$ using Equation \ref{eq:Illingworth}. It is clear that using this equation underestimates the relative error on the mass. As we have discussed, the uncertainty in the mass within $\rlimit$ is dominated by the velocity anisotropy, which is not accounted for in the M$_{\rm Ilw}$ equation, as it was derived under the assumption of isotropy. The right panel of Figure \ref{fig:Mlim} shows a comparison between the $\log_{10}$ mass error in both cases.

In conclusion, the Illingworth approximation, which was derived to only be applied on globular clusters, is a very poor estimate of the mass and mass uncertainty for dSph galaxies.

\subsection{Spitzer Formula}
\label{subsec:Spitzer}
In this subsection we slightly modify the mass estimator presented in \citet{Spitzer_69}, by halving 
their total mass, to better compare to our mass estimator:
\beq
{\rm M^{S69}_{1/2}} = 3.75 \, G^{-1} \, \avelos \, \rhalf.
\eeq
Despite the fact that this equation was derived by analyzing polytropes with indices between 
n=3 and n=5, which is a very restrictive class of mass densities that describe multi-component 
galaxies, our coefficient in Equation \ref{eq:main} is only 20\% under the Spitzer coefficient 
of 3.75. \citet{Lokas_01} find coefficients in much better agreement with ours when analyzing a 
wide variety of NFW halos, which better represent the mass density of real galaxies.

%
\begin{figure*}
\includegraphics[width=85mm]{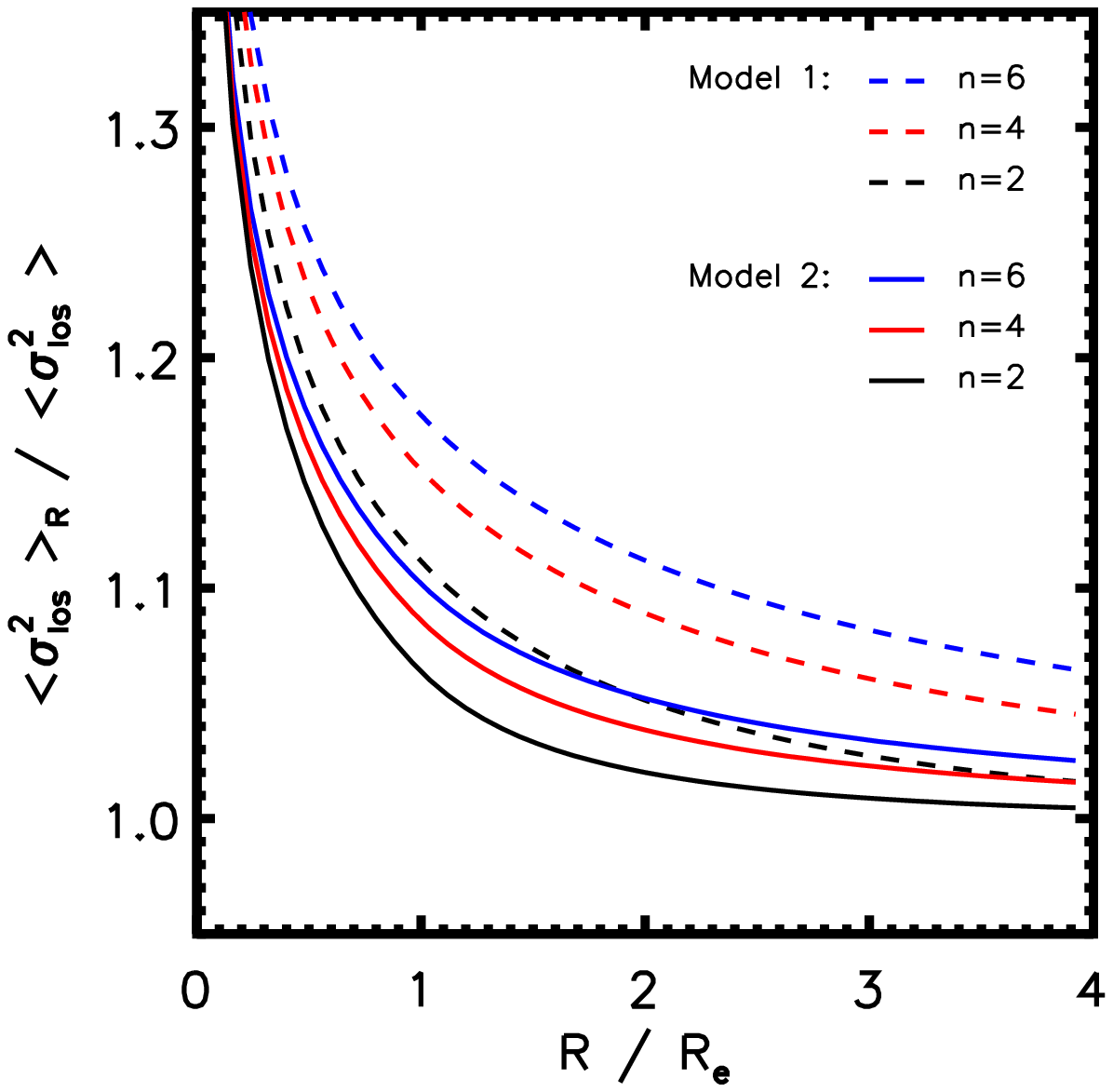}
\hspace{5mm}
\includegraphics[width=81.25mm]{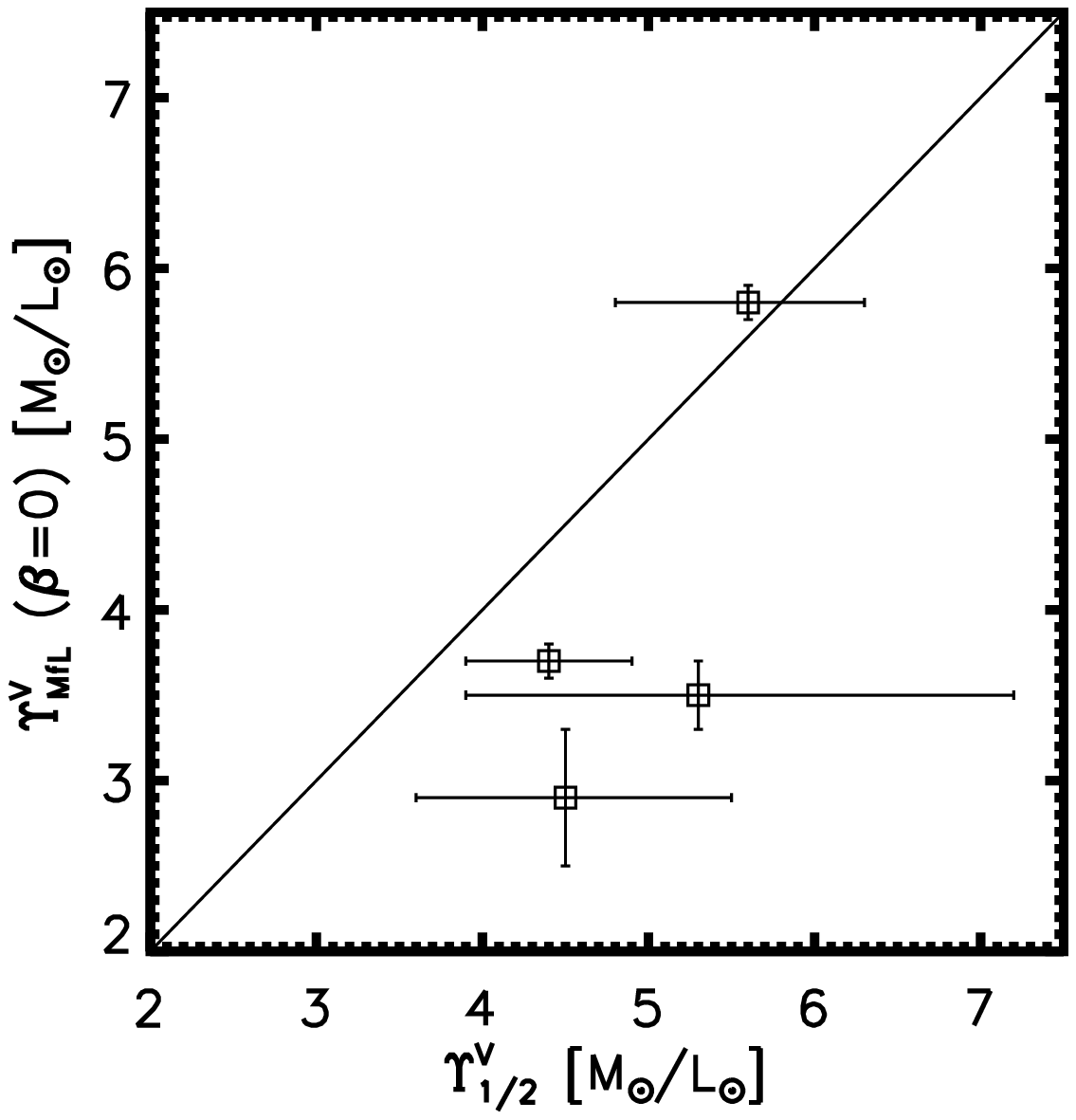}
\caption{{\em Left:} Ratio of the luminosity-weighted square of the dispersion within projected radius R divided by 
the luminosity-weighted square of the dispersion integrated to infinity for two different dispersion
models. The dashed lines (model 1) are derived by considering the median dispersion profile 
model presented in Equation 1 of \citet{Cappellari_06}: $\sigmalos (R) \propto R^{-0.066}$.
The solid lines (model 2) use the same relation to within only one effective radius. After 
$R=\Rhalf$, the dispersion profile is assumed to be flat as a function of projected radius.
The three projected S\'ersic surface brightness profiles modeled are, from top to bottom,
n=6 (blue), n=4 (red), and n=2 (black).
{\em Right:} Comparison of the derived V-band mass-to-light ratios derived using our general two-component
spherical Jeans models (x-axis) compared to those obtained under the assumption that mass follows light 
and $\beta = 0$ (y-axis). The solid line represents the one-to-one relation. The four 
galaxies modeled, from top to bottom, are NGC 4478, NGC 731, NGC 185, and NGC 855.
}
\label{fig:Cappellari}
\end{figure*}
%

\subsection{Cappellari et al. Dynamical Mass-to-Light Ratio}
\label{subsec:C06}
Using axisymmetric Jeans modeling, \citet{Cappellari_06} (hereafter C06) empirically find 
the following relation assuming a single-component mass-follows-light (MfL) density distribution:
\beq
\label{eq:C06}
{\rm \frac{M}{L}} = \frac{5 \, \avelosap \, \Rhalf}{G \, {\rm L}} \Rightarrow \Upsilon^{\rm C06}_{1/2} 
= \frac{2.5 \, \avelosap \, \Rhalf}{G \, \Lhalf} \, , 
\eeq
where $\avelosap$ is the luminosity-weighted square of the line-of-sight dispersion 
within $\Rhalf$. In practice, C06 determined $\avelosap$ by extrapolating 
the measured luminosity-weighted square of the dispersion 
within the observational aperture ${\rm R_{ap}}$. The physical aperture size varies depending
on the system, but it typically corresponds to ${\rm R_{ap}} \simeq 0.7 \, \Rhalf$ 
for the data that C06 analyzed. We continue the convention presented within this 
paper, where $\Upsilon_{1/2}$ is the dynamical mass-to-light ratio within the 3D 
deprojected half-light radius $\rhalf$. Let us rewrite our Equation \ref{eq:main} in 
order to facilitate comparison with the C06 relation:
\beq
\label{eq:W10}
\Upsilon^{\rm W10}_{1/2} = \frac{4 \, \avelos \, \Rhalf}{G \, \Lhalf},
\eeq
where we remind the reader that $\avelos$ is the luminosity-weighted square of the 
line-of-sight dispersion over the entire galaxy. In the limit that the observed velocity dispersion 
profile is perfectly flat at all radii we would expect $\avelosap = \avelos$. In this 
case the C06 estimator (with a coefficient of 2.5) is smaller by 
$\sim 40 \%$ compared to ours (with a coefficient of 4).

We explore three possible reasons for this difference in the coefficients. First, 
our analysis is explicitly spherical while the C06 models are axisymmetric. 
C06 addresses this concern by comparing their axisymmetric model results 
to spherical Jeans model results under the assumption that the velocity dispersion is
isotropic ($\beta = 0$). In this comparison, they find little difference in their dynamical
mass-to-light ratios. While this is reassuring, it is not entirely general given the assumption
of $\beta=0$ in their comparison. It remains to be seen if the geometric freedom
becomes important in comparison to more general spherical models with variable $\beta$.
In principle, projection effects can add an additional $\sim 20-30 \%$ uncertainty to spherical 
mass estimates as discussed, for example, by \citet{Gavazzi_05}. However, it would be 
surprising if these effects were systematic in biasing mass estimates.

A second possible reason for the difference in our coefficients is that $\avelosap \ne \avelos$.
Indeed, the median dispersion profile studied by C06 falls with
projected radius as $\sigma_{\rm los} (R) \propto R^{-0.066}$ over the range probed by their data
such that we would expect $\avelosap > \avelos$. In the left panel of Figure \ref{fig:Cappellari} 
we plot the ratio of the luminosity weighted square of the velocity 
dispersion as measured within an aperture radius $R$ ($\avelos_{_R}$) to the total 
luminosity-weighted square of the velocity dispersion ($\avelos$) as a function of $R$ for two 
models of $\sigma_{\rm los}(R)$ velocity dispersion profiles and several light profiles 
(S\'ersic profiles with n = 2, 4, and 6). Model 1 curves (dashed) assume
the median C06 power-law for $\sigma_{\rm los} (R)$ extends to all $R$. Model 2 curves (solid) assume
the median C06 power-law for $\sigma_{\rm los} (R)$ until $R = \Rhalf$, and then a flat  
dispersion profile for larger radii. This modification of the C06 model is motivated 
by the behavior of dispersion profiles of galaxies seen in high quality kinematics that extend
out to several effective radii \citep[e.g.,][]{Proctor_09, Weijmans_09, Geha_10}. As 
can be seen in Figure \ref{fig:Cappellari}, we expect $\avelosap / \avelos \simeq 1.1-1.2$ for the typical
aperture size (${\rm R_{ap}} \simeq 0.7 \, \Rhalf$) in the data that 
C06 analyzed. This result allows us to approximate the C06 formula as 
$\Upsilon^{\rm C06}_{1/2} \simeq {2.8 \, \avelos \, \Rhalf}/({G \, \Lhalf})$, bringing the ratio of the
C06 coefficient and our coefficient to within $\sim 30 \%$.

A third difference between our method and that of C06 is that we have allowed the dark matter
mass profile to be distinct from the light profile, while C06 assume that mass follows light. In principle, 
the MfL assumption can impose a bias because we expect the dynamical mass-to-light
ratio to increase with radius. In the right panel of Figure \ref{fig:Cappellari} we explore this issue
by comparing the dynamical V-band mass-to-light ratios of four galaxies derived using our general 
methodology to those derived under the assumption of MfL 
and $\beta = 0$. This MfL model mirrors that shown by C06 to reproduce their axisymmetric results.  
The four galaxies modeled (from top to bottom: NGC 4478, NGC 731, NGC 185, and NGC 855) were chosen as 
they had the lowest $\Upsilon^{\rm V}_{1/2}$ values in Table 1. We see that two of the galaxies have median MfL 
mass-to-light ratios that are lower by $\sim 35 \%$ than those derived for the general 
spherical case.\footnote{The reason for the abscissa errors being larger than the ordinate errors is related to the 
additional freedom that we allow in our modeling, particularly with regard to $\beta$, as we discuss 
in Section \ref{sec:mass}.} The other two galaxies do not show large differences.\footnote{We note that the 
$\Upsilon_{1/2}$ values derived from an anisotropic MfL model agree well with 
those in Table 1, as expected from Equation \ref{eq:main}.} Thus, it is possible that the MfL assumption 
can give rise to biases as large as 30\%, even in systems that are not dark matter dominated.

Future investigations that allow for non-spherical, multicomponent mass models will be 
important for investigating the advantages and limitations of the current set of assumptions
that are often used in Jeans analyses.


\end{document}